\documentclass{pas}
\usepackage{aao-macros}
\usepackage{multirow}
\usepackage{orcidlink}
\usepackage{nicematrix}
\usepackage{xspace}

\usepackage{chngcntr}

\newcommand{\fraction}[3]{$#1_{-#2}^{+#3}$\%} 
\begin{document}

\lefttitle{Publications of the Astronomical Society of Australia}
\righttitle{Oğuzhan Çakır}

\jnlPage{1}{22}
\jnlDoiYr{2026}
\doival{10.1017/pasa.xxxx.xx}

\articletitt{Research Paper}

\title{\textbf{Hector Galaxy Survey: \textit{Falling in Between} - Infalling Galaxies in the Midst of the Abell 3667 Merger}}

\author{\gn{Oğuzhan} \sn{Çakır}$^{1,2,3}$ \orcidlink{0000-0002-1045-2559}, \gn{Matt} \sn{Owers}$^{1,2,3}$ \orcidlink{0000-0002-2879-1663}, \gn{Gabriella} \sn{Quattropani}$^{1,2,3}$ \orcidlink{0009-0009-9074-716X}, \gn{Mina} \sn{Pak}$^{1,3,4}$ \orcidlink{0000-0002-5896-0034}, \gn{Stefania} \sn{Barsanti}$^{5}$ \orcidlink{0000-0002-9332-5386}, \gn{Joss} \sn{Bland-Hawthorn}$^{3,5}$ \orcidlink{0000-0001-7516-4016}, \gn{Julia} \sn{Bryant}$^{5,6}$ \orcidlink{0000-0003-1627-9301}, \gn{Scott} \sn{Croom}$^{5}$ \orcidlink{0000-0003-2880-9197}, \gn{Pratyush Kumar} \sn{Das}$^{7}$ \orcidlink{0000-0002-4326-8598}, \gn{Caroline} \sn{Foster}$^{8}$ \orcidlink{0000-0003-0247-1204}, \gn{Madusha} \sn{Gunawardhana}$^{3,5,9}$ \orcidlink{0000-0002-7301-461X}, \gn{Hyunjin} \sn{Jeong}$^{4}$ \orcidlink{0000-0002-0145-9556}, \gn{Jong Chul} \sn{Lee}$^{4}$, \gn{Joon Hyeop} \sn{Lee}$^{4}$ \orcidlink{0000-0003-3451-0925}, \gn{Yifan} \sn{Mai}$^{2,3,5,13}$ \orcidlink{0000-0003-3514-6280}, \gn{Kyuseok} \sn{Oh}$^{4}$ \orcidlink{0000-0002-5037-951X}, \gn{Sree} \sn{Oh}$^{10}$ \orcidlink{0000-0002-4731-9604}, \gn{Andrei} \sn{Ristea}$^{11,12}$ \orcidlink{0000-0003-2723-0810}, \gn{Sarah} \sn{Sweet}$^{7}$ \orcidlink{0000-0002-1576-2505}, \gn{Sujeeporn} \sn{Tuntipong}$^{5}$ \orcidlink{0009-0002-8534-5077}, \gn{Gurashish} \sn{Bhatia}$^{6}$, \gn{David} \sn{Brodrick}$^{9}$, \gn{Rebecca} \sn{Brown}$^{6,14}$, \gn{Elton} \sn{Cheng}$^{6}$, \gn{Robert} \sn{Content}$^{14}$, \gn{Fred} \sn{Crous}$^{6}$, \gn{Tony} \sn{Farrell}$^{14}$, \gn{Peter} \sn{Gillingham}$^{14}$, \gn{Ellen} \sn{Houston}$^{14}$, \gn{Jon} \sn{Lawrence}$^{14}$, \gn{Helen} \sn{McGregor}$^{14}$, \gn{Mahesh} \sn{Mohanan}$^{14}$, \gn{Seong-Sik} \sn{Min}$^{6}$, \gn{Barnaby} \sn{Norris}$^{5,6}$, \gn{Naveen} \sn{Pai}$^{14}$, \gn{Ayoan} \sn{Sadman}$^{6}$, \gn{Will} \sn{Saunders}$^{14}$, \gn{Sudarshan} \sn{Venkatesan}$^{14}$, \gn{Adeline} \sn{Wang}$^{6}$, \gn{Ross} \sn{Zhelem}$^{14}$ and \gn{Jessica} \sn{Zheng}$^{14}$}


\affil{$^1$School of Mathematical and Physical Sciences, Macquarie University, Sydney, NSW 2109, Australia,
$^{2}$Astrophysics and Space Technologies Research Centre, Macquarie University, Sydney, NSW 2109, Australia,
$^{3}$ARC Centre of Excellence for All-Sky Astrophysics in 3 Dimensions (ASTRO-3D),
$^{4}$Korea Astronomy and Space Science Institute (KASI), 776 Daeduk-daero, Yuseong-gu, Daejeon  34055, Republic of Korea,
$^{5}$Sydney Institute for Astronomy (SIfA), School of Physics, The University of Sydney, NSW 2006, Australia,
$^{6}$Astralis-USydney, Sydney Institute for Astronomy (SIfA), School of Physics, The University of Sydney, NSW, 2006, Australia,
$^{7}$School of Mathematics and Physics, University of Queensland, Brisbane, QLD 4072, Australia,
$^{8}$School of Physics, University of New South Wales, Sydney, NSW 2052, Australia,
$^{9}$Research School of Astronomy and Astrophysics, Australian National University, Canberra, ACT 2611, Australia
$^{10}$Department of Astronomy and Yonsei University Observatory, Yonsei University, Seoul 03722, Republic of Korea,
$^{11}$Centre for Astrophysics and Supercomputing, Swinburne University of Technology, John St, Hawthorn, 3122, Victoria, Australia,
$^{12}$ARC Centre of Excellence in Optical Microcombs for Breakthrough Science (COMBS), Australia,
$^{13}$Australian Astronomical Optics, Macquarie University, Sydney, NSW 2109, Australia,
$^{14}$Astralis-AAO, Australian Astronomical Optics, Faculty of Science and Engineering, Macquarie University, NSW 2109, Australia}

\corresp{O. Çakır, Email: oguzhan.cakir@students.mq.edu.au} 

\citeauth{Oğuzhan Ç. et al., Hector Galaxy Survey: \textit{Falling in Between} - Infalling Galaxies in the Midst of the A3667 Merger, {\it Publications of the Astronomical Society of Australia} {\bf 00}, 1--22. https://doi.org/10.1017/pasa.xxxx.xx}

\history{(Received xx xx xxxx; revised xx xx xxxx; accepted xx xx xxxx)}

\begin{abstract}
Whether cluster mergers enhance ram pressure stripping (RPS) and accelerate the subsequent evolution of member galaxies remains an open question. In this paper, we address this question by investigating galaxy populations in the nearby merging cluster Abell 3667 ($z\simeq0.0553$) using spatially resolved data from the Hector Galaxy Survey. We define an RPS sample by combining Hector-selected galaxies with ionised gas disturbances (e.g., asymmetric tails or truncated disks) and supplementary, optically identified jellyfish galaxies lacking Hector data. The majority of the RPS sample ($\sim$\fraction{71}{10}{7}; 20/28) lies within $\rm R_{200}$, where the impact of the merger is greater. Most asymmetric galaxies ($\sim$\fraction{73}{14}{8}; 11/15), especially those with extreme RPS signatures, are concentrated towards the inner cluster regions ($\rm R\lesssim0.6\ R_{200}$), distributed along the merger axis between two shock-tracing radio relics. These central asymmetric galaxies present two separate spatial and kinematic concentrations: a population located at the North-West (NW) subcluster and downstream of its radio relic, a region of high-velocity intracluster medium (ICM) bulk motion, with blueshifted line-of-sight velocities, and a primarily redshifted population close to the main cluster (MC), which also shows evidence of a turbulent ICM. Despite their projected spatial association with the MC core and the NW substructure, the velocities of both samples indicate they are not bound to these merging substructures. Tail orientations provide additional insight into their orbital histories: NW tails point away from the cluster centre and are frequently aligned with the merger axis, suggestive of merger-driven stripping, while MC tails show neither pattern clearly. The tails are also broadly westward-pointing, with the MC tails closely tracing due west and the NW tails shifted towards the northwest, pointing to two distinct filamentary accretion events for the NW and MC populations. Taken together, our results indicate enhanced RPS activity in the heart of A3667, primarily driven by infalling galaxies accreted along nearby filaments and interacting with the merger-driven turbulent environment.
\end{abstract}


\maketitle

\section{Introduction}
A major merger between two equal-mass clusters is the most energetic event ($\rm\sim \!10^{63-64}\ erg$) in the universe since the Big Bang \citep{Markevitch1998}. Since roughly 25–40\% of galaxy clusters at $z < 1$ exhibit ongoing dynamical activity \citep{Geller1982, JF1999, Mann2012}, studying cluster mergers is essential to determine how these extreme environments drive galaxy evolution.
\\
\\
Galaxy clusters curtail star formation activity in the constituent galaxies, notably through gas stripping \citep{Boselli2006, Cortese2021, Boselli2022}. As galaxies fall into a cluster, their gas reservoir can be stripped by the hot ($\rm T\sim10^{7}\ K$), diffuse ($\rm 10^{-3}\ particles \ cm^{-3}$) intracluster medium \citep[ICM;][]{GG1972}, the process known as ram pressure stripping \citep[RPS;][]{GG1972}. Gas stripping may occur over a longer timescale, with the hot halo being removed, leading to a gradual decline in star formation \citep[strangulation; ][]{LTC1980}, or more rapidly by being exposed to stronger RPS, which can even remove the cold gas component embedded in the stellar disk \citep{Roediger2006, Kapferer2009}. Stronger RPS is also predicted to trigger star formation within the galaxy by compressing the cold gas \citep{Fujita1999b, Bekki2003} or ram-pressure-driven mass flows to the galaxy centre \citep{Bekki1999, Kronberger2008a, Zhu2024}.
\\
\\
RPS leaves multi-wavelength signatures, including truncated gas disks and extra-planar gas; in extreme cases, it manifests as ``jellyfish" galaxies with long stripped gas tails that host in-situ star formation \citep{Kenney1999, Gavazzi2001, Kenney2004, Cortese2007, Bekki2009, Chung2009, Smith2010, Sun2010, Owers2012, Ebeling2014, Poggianti2016, Owers2019, Stroe2020, Roberts2021a, Poggianti2025, Sun2025, Edler2026, Cakir2026}. These signatures have been commonly used to identify RPS-affected galaxies. Two main approaches have been adopted for identification, either by visual classification of multi-wavelength imaging \citep{Poggianti2016, Yoon2017} or through structural measurements such as concentration and asymmetry \citep{McPartland2016, Roberts2021b, Krabbe2024, Cakir2026}. In a recent study, \citet{Cakir2026} systematically searched for RPS-affected galaxies using spatially resolved data from the Sydney-AAO Multi-object Integral-field unit (IFU) spectrograph \citep[SAMI;][]{Croom2012} Galaxy Survey \citep[SAMI-GS;][]{Bryant2015}. They visually inspected the $\rm H\alpha + [NII]\lambda6584$ emission maps to identify galaxies with gas asymmetries (asymmetric galaxies) and truncated emission (truncated galaxies), as an indicator of recent or ongoing RPS. They found that asymmetric galaxies are found close to the cluster centre ($\rm R\lesssim 0.5\ R_{200}$, where $\rm R_{200}$ is the characteristic cluster radius enclosing the region with the density exceeding 200 times the critical density), with higher peculiar velocities ($\mu(|v_{pec}|)=1.15\ \sigma_{200}$, where $v_{pec} = c \times (z_{galaxy}-z_{cluster}) /(1+z_{cluster})$, with $c$ being the speed of light), as expected from an infalling population, unlike their truncated counterparts with lower velocities ($\mu(|v_{pec}|)\sim0.76\ \sigma_{200}$) and being located at larger cluster-centric distances ($\rm R\gtrsim 0.5\ R_{200}$). They concluded that asymmetric galaxies are currently undergoing RPS, while truncated galaxies represent a post-RPS population.
\\
\\
Cluster mergers host more violent environments with respect to their virialised counterparts. In the course of the core passage, the merging haloes can travel extremely fast \citep[e.g., relative speeds up to $\rm \sim\!4500 \ km\ s^{-1}$;][]{Sarazin2002, Markevitch2002, Markevitch2004, Owers2011, Owers2014}. This generates supersonic shocks traversing the ICM \citep{Ryu2003, Markevitch2007, Lee2025}, leading to a pressure jump in the ICM. Hence, merging clusters are expected to enhance RPS activity, observationally supported by previous studies reporting elevated RPS activity in cluster mergers. Numerical simulations showed both an increase in star formation \citep{Bekki2010, Roediger2014, Ruggiero2019, Aldas2025} and a decline due to increased gas loss \citep{Fujita1999a, Domainko2006, Li2023} during the major mergers. Observationally, this dichotomy still exists, where an increase in star formation is found \citep{Rawle2014, Stroe2017, Roman-Oliveira2019}, while other studies support a rapid quenching \citep{Ma2010, Pranger2014, Deshev2017, Roberts2024}. 

The enhanced star formation activity is often spatially coincident with cluster regions where merger-related features, such as substructures and shocks, are present. Using Hubble Space Telescope (HST) imagery, \citet{Owers2012} identified four jellyfish galaxies in A2744 \citep[$z = 0.3064$;][]{Owers2011}, with three of them being found in the vicinity of a merging substructure and its shock front detected in X-ray. In the same cluster, \citet{Bellhouse2022} investigated a jellyfish sample, partially covering those in \citet{Owers2012}, using $\rm H\alpha$ emission mapping detected with the Multi Unit Spectroscopic Explorer \citep[MUSE; ][]{Bacon2010} instrument, and found that all RPS galaxies are associated with the high-velocity blueshifted substructure, indicating merger-induced activity. 

Similarly, investigating the galaxy population in a young ($\sim$1 Gyr post-core passage) merging cluster at $z\sim0.2$, \citet{Stroe2015} reported a high incidence of $\rm H\alpha$ emitters with star formation rates (SFRs) above $\rm 0.17 \ M_\odot \ yr^{-1}$,  located in regions close to the merger shocks. This enhancement yielded a cluster-wide SFR density of $\rm 6.97 \  M_\odot \ yr^{-1} \ Mpc^{-3}$, comparable to that observed in the field galaxies at cosmic noon ($z\sim2$). As a follow-up, \citet{Stroe2020} investigated five $\rm H\alpha$ emitters near the shocks, using spatially resolved data from Gemini Multi-Object Spectrograph (GMOS) IFU spectroscopy. They found that all galaxies exhibit disturbance in their ionised gas aligned with the merger axis, confirming shock-induced interaction.
 
Broader statistical analyses also confirm these trends across larger cluster samples. In this context, \citet{McPartland2016} examined 223 RPS candidates identified within a massive cluster sample at $z=0.3-0.7$. Combining the cluster dynamical states, they found that the extreme RPS events (i.e., jellyfish galaxies) preferentially occur in cluster mergers. At lower redshifts ($z<0.1$), \citet{Lourenco2023} conducted a homogeneous analysis on RPS candidates from the GAs Stripping Phenomena in galaxies with the MUSE \citep[GASP;][]{Poggianti2016} survey, finding that interacting clusters exhibit a mild enhancement in the frequency of stripping candidates.

Offering an alternative explanation, \citet{Cakir2025} studied star formation activity across a subsample of clusters from the SAMI-GS, with varying merger stages. Using single-fibre spectroscopy from the SAMI Cluster Redshift Survey \citep{Owers2017} and Sloan Digital Sky Survey-III \citep{Alam2015}, they used star-forming galaxy fractions as a proxy of ongoing star formation activity. They found that merging clusters host more ($\sim$29\%) star-forming galaxies within $\rm R_{200}$---where the impact of the merger is pronounced---relative to relaxed clusters ($\sim$24\%). However, they found no enhancement in $\rm H\alpha$ line strengths linked to triggered star formation activity. They instead suggested that the central star-forming population consists of a projection of unquenched member galaxies and infalling galaxies accreted from nearby filaments, rather than galaxies with merger-elevated activity. This picture is consistent with the spatial mixing/redistribution scenarios detailed in \citet{Fujita1999a} and \citet{Cohen2014, Cohen2015} for actively assembling clusters.
\\
\\
The tails observed in RPS candidates help us to constrain their orbital motion. While tails typically point away from the cluster centre in virialised systems \citep{Smith2010, RP2020, Liu2021, Roberts2021b, Kolcu2022, Smith2022, George2024}, suggesting radial infall of galaxies observed prior to pericentre, this trend is often weaker or absent in interacting clusters \citep{Roman-Oliveira2019, Salinas2024}. For example, in the post-merger cluster A1758N, \citet{Ebeling2019} found that the majority of RPS tails were oriented towards the SE, suggesting a bulk flow originating from a nearby filament rather than isotropic infall from the field. Furthermore, several investigations of merging clusters report a clear alignment between ionised gas disturbances and the merger axis \citep{Rawle2014, Stroe2020, Edler2026}, providing direct evidence for merger-driven evolution. Taken together, such misalignments between gas tails and the cluster centre or merger axis are, therefore, useful probes to understand the merger geometry/activity and to distinguish galaxy populations with orbital history.
\\
\\
Although systematic searches for RPS candidates using spatially resolved data have been conducted in typical cluster environments \citep{Cakir2026}, IFU studies targeting the direct link between RPS and the intense dynamical activity of extreme major mergers, expected to further enhance RPS as outlined above, remain highly limited in the literature. In this context, we use the new Hector Galaxy Survey \citep[Hector-GS;][]{Bryant2024, Oh2025}, which has been observing a large sample of low-redshift galaxies across different environments, including 11 massive galaxy clusters with varying dynamical activity, some of which exhibit extreme major mergers. Moreover, the Hector-GS is specifically designed to sample galaxies in the outskirts out to $\rm 2\ R_{200}$---regions where both infall and pre-processing are expected to play critical roles in galaxy evolution, yet are frequently undersampled by previous IFU studies. Therefore, the Hector-GS will enable us to study the impact of the dynamical state of a cluster on galaxy evolution by providing a complete picture out to the cluster outskirts. In this early science paper, we focus on the nearby Abell 3667 cluster, which is undergoing an extreme major merger \citep{Rottgering1997, Owers2009, deGasperin2022, Omiya2024}, as a case study to search for direct evidence of potential merger-induced RPS activity.
\\
\\
The paper is organised as follows: Section~\ref{sec:Data and Sample} introduces the Hector-GS, data products and the sample used in this study. In Section~\ref{sec:Analysis and Results}, we present our analysis and results on spatial distribution (\S\ref{subsec:Spatial Distribution}), the peculiar velocities (\S\ref{subsec:Velocity Distribution}), and tail orientations (\S\ref{subsec:Tail Orientations}). Section~\ref{sec:Discussion} interprets our findings by comparing with the existing literature, and Section~\ref{sec:Conclusion} highlight the main findings in this paper. Throughout this paper, a flat $\Lambda$CDM cosmology is assumed with $\mathrm{H_0 = 70 \ km \ s^{-1} \ Mpc^{-1}}$, $\Omega_\mathrm{m} = 0.3, \ \Omega_\Lambda = 0.7$. 

\section{Data and Sample}\label{sec:Data and Sample}
\subsection{Hector Galaxy Survey}
The Hector-GS \citep{Bryant2024, Oh2025} is an integral field spectroscopy (IFS) project being carried out on the 3.9 m Anglo-Australian Telescope (AAT), succeeding the SAMI Galaxy Survey \citep{Croom2012, Bryant2015}. The Hector-GS is aiming to observe up to 15,000 low-redshift ($z<0.1$) galaxies to decipher the diversity of galaxies across different environments. Overall, galaxies targeted in the Hector-GS have redshifts between $0<z<0.1$ and stellar masses $\rm 7< log(M_\ast/M_\odot)<12$. Details of the target selection for the Hector-GS will be provided in an upcoming paper (Barsanti et al., in preparation). As of the internal Hector Data Release (DR; December 2024), the survey comprises 2375 observations (including duplicates) for 1664 unique galaxies. Of these, 934 correspond to 625 unique cluster members, with membership allocations provided by Owers et al. (in preparation).

The Hector instrument \citep{Bryant2024} is mounted at the prime focus of the AAT, utilising 19 fused fibre bundles \citep[hexabundles;][]{Bland-Hawthorn2011, Bryant2014, Brown2018, Wang2019, Wang2020, Wang2023} with a high (75\%) fill factor, spread over a 2-degree field of view (FoV). Hector's large FoV is the most important aspect for cluster science by allowing us to observe galaxies in the central $\rm R_{200}$ and also the outskirts ($\rm 1-2\ R_{200}$) in one exposure. Compared to SAMI, Hector hexabundles vary in size, containing between 61 and 169 fibres of 1.6-arcsecond diameter, which results in IFU fields of view ranging from 15 to 26 arcseconds. At the redshift of Abell 3667 ($z \sim\!0.0553$; \citealt{Owers2009},  physical scale of $1.074\ \rm kpc/^{\prime\prime}$), these apertures correspond to a physical diameter span of approximately 16 to 28\ $\rm kpc$. This spatial coverage allows galaxies to be observed out to 2 effective radii ($R_{\rm e}$) for $\sim70\%$ of the sample, providing sufficient footprint to capture asymmetric features within the aperture. Hector also utilises two additional 37-core hexabundles allocated to secondary standard stars for calibration purposes. The hexabundles feed into either the double-beam AAOmega spectrograph \citep{Sharp2006} or the Hector spectrograph ``Spector" \citep{Content2018, Zhelem2022, Bryant2024}. Both instruments provide dual-arm coverage across the optical regime (3750--7800\,\AA) with a spectral resolution of $R \sim\!1800$--$5600$; full configuration, resolution, and line spread function details are comprehensively described in \citet{Bryant2024} and \citet{Oh2025}.
\\
\\
The Hector-GS has been observing galaxies selected from the Wide Area VISTA Extra-Galactic Survey \citep[WAVES;][]{WAVES} regions and eleven massive galaxy cluster regions ($\sigma_{200}>650 \ \mathrm{km \ s^{-1}}$; \citealt{Biviano2017, Owers2017}, corresponding to $\rm log\ M_{200}/M_\odot\gtrsim 14.5$ using the scaling relation given in \citealt{Munari2013}). Eight of the eleven clusters are selected from the OmegaWINGS survey \citep[A151, A3158, A3266, A3376, A3391/95, A3667, and A3716;][]{Moretti2017, Biviano2017} and three from the SAMI-GS \citep[A85, A119, A2399;][]{Owers2017}. The eleven clusters have been the subject of an extensive redshift survey that extends coverage to $\sim\!2.5\ \mathrm{R_{200}}$ at a limiting magnitude $r=19.5$ with high spectroscopic completeness ($\sim\!90\%$; Owers et al., in preparation). 

For the analysis presented in this paper, cluster membership for Abell 3667 (as well as the broader Hector cluster sample) is determined through an iterative, multi-step process outlined in \citet{Owers2017} based on galaxy positions and redshifts. Initially, obvious interlopers are removed by excluding galaxies with $|v_{pec}| > 5000 \ \mathrm{km \ s^{-1}}$ and $\rm R>6 \ Mpc$. The remaining galaxies are used to estimate the cluster's $\rm R_{200}$ and $\sigma_{200}$ iteratively. Once $\sigma_{200}$ is determined, a secondary kinematic cut is applied at $|v_{pec}| > 3.5\sigma_{200}$. The next step involves stricter interloper rejection by identifying gaps in the velocity distribution at different cluster-centric distances using a similar approach to ``shifting-gapper" \citep{Fadda1996}. Final cluster membership is determined by including only those galaxies within the escape velocity envelope that is defined in projected phase space using the caustics technique \citep{Diaferio1999}.

\subsection{Products}
\subsubsection{Emission Line Measurements}
The fitting procedure for emission line properties will be described in Quattropani et al. (in preparation), which follows the same procedure outlined in \citet{Owers2019}. Briefly, the first step is to model the underlying continuum, requiring penalized Pixel Fitting \citep[\texttt{pPXF},][]{Cappellari2004, Cappellari2017} to be run multiple times, as follows:
\begin{itemize}
    \item An initial \texttt{pPXF} run is performed on both per-spaxel and binned spectra with a 12$\rm ^{th}$-order additive polynomial as in \citet{Sande2017} to derive stellar and gas kinematics.
    \item The second \texttt{pPXF} fit is performed on the binned spectrum, holding stellar and gas kinematics fixed and using a 12$\rm ^{th}$-order multiplicative polynomial.
    \item The third and final \texttt{pPXF} fit is applied to individual spaxels using a 12$\rm ^{th}$-order multiplicative polynomial. If the signal-to-noise ratio (SNR) is less than 5, the weights of the stellar templates from the previous step are used to generate an optimal template. Otherwise, a combination of the non-zero weighted templates is used.
\end{itemize}
Once the stellar continuum is modelled, it is subtracted from the raw spectrum to yield an emission-only spectrum. Multiple Gaussians, up to three components, are fit to the emission spectrum using a method similar to that of the \texttt{LZIFU} code \citep{Ho2016}. The number of final components is then determined through the Bayesian information criterion (BIC). For our analysis, we use $\rm H\alpha$, $\rm [NII]\lambda6584$, $\rm H\beta$, and $\rm [OIII]\lambda5007$ emission lines. Here, we only consider 1-component fits to include most of the flux.

\subsubsection{Spectroscopic Classification}\label{subsection:Spectroscopic Classification}
We adopt the classification scheme described in Tables 1 and 2 in \citet{Owers2019}, in which the emission and absorption line features are treated simultaneously. A summary of the classification procedure is given below.
\\
\\
Emission-line classification is performed on spectra exhibiting $\rm H\alpha \text{ and/or } [NII]\lambda 6584$ as the primary lines, alongside $\rm H\beta \text{ and/or } [OIII]\lambda 5007$ lines. We require reliable detections with a signal-to-noise ratio (SNR) $>$ 3, where $\rm SNR=flux/\sigma_{flux}$. For the primary lines, we additionally seek an equivalent width (EW) $>$ 1 \AA, to avoid mismatches between templates and spurious detections. Depending on the availability of the secondary lines, we utilise the Baldwin-Phillips \& Terlevich \citep[BPT;][]{BPT1981} or WHAN \citep{CidFernandes2010} diagnostics to classify the ionising sources. Ionisation sources are divided into star-forming (SF), intermediate (INT), and non-star-forming (NSF), following the demarcation lines from \citet{Kewley2001} and \citet{Kauffmann2003}. In cases where only one primary line is detected with a secondary line, the spectrum is SF if $\rm H\alpha$ is present, or NSF if $\rm [NII]\lambda 6584$ is detected. Moreover, based on the $\rm H\alpha$-line strength, EW($\rm H\alpha$), we further sub-classify these main classes into 
\begin{itemize}
    \item retired (r) if $\rm EW(H\alpha)/\AA<3$ for INT and NSF,
    \item weak (w) if $\rm EW(H\alpha)/\AA<3$ for SF, and $\rm3<EW(H\alpha)/\AA<6$ for NSF,
    \item strong (s) if $\rm EW( H\alpha)>6\ \AA$ for only NSF
\end{itemize}
\noindent which are defined by \citet{CidFernandes2010}.
\\
\\
Absorption line classification is performed on the spectra labelled as passive (that do not present at least two emission lines) or those whose emission originates from non-star-forming ionisation. We measure the EWs of $\rm H\delta, H\gamma, H\beta$ absorption lines following the procedure in \citet{Owers2019}, using emission-subtracted spectra, when the emission is reliable. In order to increase the $\rm H\delta$ absorption signal, employing the linear relations given in Figure 3 of \citet{Owers2019}, we generate pseudo-EW($\rm H\delta$) from the EWs of $\rm H\beta$ and $\rm H\gamma$. We then calculate the weighted average of $\rm EW(H\delta)$ and the pseudo-EW measurements to obtain EW($\rm \overline{H_{\delta\gamma\beta}}$), using the inverse variances as weights. For the spectra with $\rm SNR(4100\ \AA) > 3\ \AA^{-1}$, we classify those with $\rm SNR(EW(\overline{H_{\delta\gamma\beta}}))>3$, where $\rm SNR(EW(\overline{H_{\delta\gamma\beta}}))=|EW(\overline{H_{\delta\gamma\beta}})| /\sigma_{EW(\overline{H_{\delta\gamma\beta}})}$, and
\begin{itemize}
    \item $\rm EW(\overline{H_{\delta\gamma\beta}})<-3\ \AA$ as $\rm H\delta-$strong (HDS),
    \item $\rm EW(\overline{H_{\delta\gamma\beta}})>-3\ \AA$ as passive.
\end{itemize}
\noindent The global classification per galaxy is assigned considering the spaxels with $\rm SNR(4100\AA)>3\ \AA^{-1}$. If more than 90\% of the spaxels are classified as passive, rNSF, wNSF, sNSF, rINT, or wSF, the galaxy is classified as a passive galaxy (PASG). In contrast, SF galaxies (SFG) and HDS galaxies (HDSG) both must have at least 10\% of their spaxels classified as SF or INT, and HDS or NSF\_HDS, respectively. For SFGs and HDSGs, an additional continuity criterion is applied, in that a spaxel to be counted within the 10\% threshold must have at least three neighbouring spaxels classified as one of the classes defined for SFG or HDSG, respectively. 

\subsection{Sample Selection}\label{sec:Sample selection}
\subsubsection{Abell 3667}
In this study, we focus on a well-known nearby merging cluster, Abell~3667 (hereafter A3667; $z \simeq 0.0553$, \citealp{Owers2009}; $\rm R_{200} \simeq 2.3$~Mpc and 
$\sigma_{200} \simeq 953\ \rm km\ s^{-1}$, Owers et al., in preparation), which is targeted for early science with the Hector-GS that provides data for 219 galaxies as detailed in the next section. Multi-wavelength studies present clear evidence for ongoing dynamical activity, with the merger occurring almost entirely within the plane of the sky \citep{Owers2009}. Radio observations reveal double radio relics, associated with merger shocks, with one to the North-West (NW) and one to the South-East (SE), a radio bridge, and a radio halo \citep{Rottgering1997, Carretti2013, deGasperin2022}. At X-ray wavelengths, previous studies \citep{Knopp1996, Vikhlinin2001, Mazzotta2002, Markevitch2007, Owers2009b, Omiya2024, Omiya2025} reveal an elongated emission, ICM plumes, and a cold front located to the southeast of the central brightest cluster galaxy (BCG), whose coordinates we adopt as the cluster centre ($\rm \alpha_{J2000} = 303.1139^\circ, \ \delta_{J2000} = -56.8268^\circ$). Optical analysis shows a bimodal spatial distribution, with a secondary peak centred on the secondary BCG \citep{Proust1988, Sodre1992, Owers2009}, which is also identified through lensing studies \citep{Joffre2000}. In a more detailed study, \citet{Owers2009} partitioned A3667 into five substructures using Kaye's Mixture Model \citep[KMM; ][]{Basford1985}, with KMM5 being the main cluster ($\overline{v} = -103\pm71 \ \rm km\ s^{-1}$, $\sigma=1073\pm64 \ \rm km\ s^{-1}$), KMM2 being the NW cluster ($\overline{v} = 422\pm82 \ \rm km\ s^{-1}$ , $\sigma=1039\pm66 \ \rm km\ s^{-1}$), and three relatively minor substructures. Taken together, the observational evidence for A3667 is consistent with an offset binary merger, with the core passage occurring $\sim1$ Gyr ago, based on simulation estimates \citep{Roettiger1999}.

\subsubsection{Best Cubes}\label{subsec:Best Cubes}
The Hector-GS DR provides data cubes for 219 galaxies in A3667. Cross-matching these with the currently available emission-line products yields 318 data cubes for 205 unique galaxies.

For the galaxies with repeated observations, we define the best cube as the one observed using a larger hexabundle. If hexabundle sizes are equal, we select the cube with the higher median $\rm SNR( 4000<\lambda_{rest}/\AA<5000)$, measured within a 5-arcsecond-diameter aperture. While defining the best cubes, we identified that 12 cubes (including four unique observations) suffer from poor primary/secondary flux calibration, which results in poorly measured continuum and, therefore, emission line properties. We exclude these cubes from further analysis, irrespective of their hexabundle size and SNR. This selection leaves us with 201 unique cubes/galaxies. 

As raised in Section 3 of \cite{Oh2025}, nearby objects (e.g., stars) within the hexabundle impacted the centring of some cubes. Three galaxies in the above selection suffer from miscentring due to contamination from nearby objects, although they were previously selected as the best cubes. Similarly, five galaxies were not able to be classified as "PASG", "SFG", or "HDSG" due to low SNR($\rm 4100 \AA$) or not satisfying the criteria described in Section~\ref{subsection:Spectroscopic Classification}. After removing all these cases, we have 193 unique cubes/galaxies as our main sample. 

\subsubsection{RPS candidates}\label{RPS candidates definition}
As the primary focus of this paper, we need to identify galaxies undergoing or having recently undergone RPS to study environment-driven evolution. The RPS candidates are primarily selected based on two criteria: the disturbances in the ionised gas morphology---asymmetries or truncation---and the presence of $\rm H\delta$-strong regions to trace recently quenched regions.
\\
\\
\textit{\textbf{\textcolor{red}{Asymmetric and Truncated Galaxies:}}} We adopt a two-step approach to identifying RPS candidates that includes a quantitative pre-selection and then visual confirmation. The pre-selection of galaxies with ionised gas asymmetries (asymmetric) and truncation (truncated) is done using morphological parameters: \textbf{\textit{i)}} shape asymmetry \citep[$\rm A_{shape}$,][]{Pawlik2016} defined as follows 
\[\rm A_{shape} = \frac{\sum{|I_0 - I_{180}|}}{2\times\sum{I_0}}\] where $\rm I_0$ is the binary detection map and $\rm I_{180}$ is the flux map rotated by $180^\circ$ around the galaxy centre, and \textbf{\textit{ii)}} ionised gas concentration \citep[$\rm C_{emission}$,][]{Schaefer2017} expressed as \[ \rm C_{emission}=\frac{r_{50}(H\alpha+[NII]\lambda 6584)}{r_{50}(Continuum)}\] with $\rm r_{50}$ being the radius enclosing half of the flux. In order to measure these parameters, we generate the ionised gas ($\rm H\alpha+[NII]\lambda6584$) maps. Following \citet{Cakir2026}, who demonstrate that a threshold of $\rm A_{shape} > 0.2$ identifies asymmetric galaxies while truncated galaxies exhibit $\rm log(C_{emission})<0$, we pre-select our asymmetric and truncated galaxy candidates according to these criteria. Additionally, because RPS does not always manifest as SF-related asymmetric features/tails \citep{Fossati2016, Owers2019, Pedrini2022, Cakir2026, Foster2026}, our asymmetric galaxy sample also intentionally includes non-SFGs. In contrast, our truncated galaxy selection is limited exclusively to SFGs. This allows us to focus specifically on the truncation of active star-forming disks, while effectively excluding galaxies with centrally concentrated emission powered by non-environmental mechanisms such as AGN and LIERs/LINERs. This pre-selection yields 50 asymmetric and 41 truncated galaxy candidates. Subsequent visual inspection of the ionised gas map overlaid onto stellar continuum counterparts confirmed 12 truncated and 12 asymmetric galaxies. False-positive candidates were rejected for the following reasons: \textbf{\textit{i)}} significant contamination of the nebular emission or stellar continuum by nearby objects, \textbf{\textit{ii)}} insufficient spatial extent of the nebular emission (detected in only a few spaxels), and \textbf{\textit{iii)}} aperture effects. The latter case emerges when the nebular emission fills the entire hexabundle, but the bundle centre is not coincident with the galaxy/cube centre, or when the hexabundle partially covers the galaxy, yielding spurious asymmetry or preventing a clear visual assessment. It is also worth noting that since the median seeing for the Hector survey is $\sim$2 arcsec \citep{Oh2025}, consistent with that of the SAMI survey \citep{Croom2021}, we do not expect any systematic differences relative to the thresholds defined by \citet{Cakir2026} that might affect the completeness of our RPS sample.
\\
\\
\textit{\textbf{\textcolor{red}{HDSGs:}}} This is simply achieved by selecting HDSGs based on their spectral features. Note that since this selection is based on absorption properties, there may be overlap between HDSGs and truncated and asymmetric galaxies. Indeed, five out of six HDSGs are classified as asymmetric galaxies based on their ionised gas morphologies. Spectroscopic classification maps shown in Figures~\ref{appendix:Asymmetric Galaxies} and \ref{appendix:HDS-q} indicate that four HDSGs host central star formation, while one (ID:\texttt{C901005402400486}) is completely quenched (hereafter HDS-q).
\\
\\
\textit{\textbf{\textcolor{red}{Jellyfish galaxies/candidates:}}} To provide a more comprehensive view of RPS in A3667, in addition to the candidates identified above, we include jellyfish galaxies yet unobserved by Hector-GS but with available optical imaging, identified either previously or newly in this study. For this purpose, we first utilise the sample from \citet{Poggianti2016}, where candidates were identified using B-band imaging from the WIde-field Nearby Galaxy-cluster Survey \citep[WINGS; ][]{Fasano2006} and OmegaWINGS \citep{Gullieuszik2015}. This provides two additional candidates within A3667: JO170 and JO171. Among these, JO171 is a well-studied galaxy with a Hoag's object-like morphology, consisting of a circular star-forming ring and an older core separated by a distinct gap \citep{Moretti2018}. It exhibits prominent star-forming tails, providing evidence of extreme RPS, with a jellyfish class \citep[\texttt{JClass}=5;][]{Poggianti2016}. In contrast, JO170 has not been studied in the same depth and was previously noted for exhibiting only a marginal tail \citep[\texttt{JClass}=1;][]{Poggianti2016, Salinas2024}. However, as shown in the right panel of Figure~\ref{fig:Jellyfish Galaxies}, deeper $griz$ imaging from the Legacy Survey Data Release 10 \citep[LS DR10;][]{Dey2019} reveals an extended tail associated with JO170, spanning $\sim65-125$ kpc in projection for our adopted cosmology, as a sign of extreme ongoing stripping. Apart from the \citet{Poggianti2016} sample, we also conduct a visual inspection of cluster members from Owers et al. (in preparation) to select a sample of jellyfish candidates based on $griz$ imaging from the LS DR10. We look for strong indications of stripping in the form of stripped star-forming tails, similar to \texttt{JClass} = 4 and 5 defined in \citet{Poggianti2016}. Only one galaxy (ID:\texttt{C901005481506069}; shown in the left panel of Figure~\ref{fig:Jellyfish Galaxies}) from LS DR10 is returned in addition to JO170 and JO171, as a strong jellyfish candidate.
\\
\\
The above selection yields a final sample of 28 RPS candidates, whose breakdown is given in Table~\ref{tab:RPS Sample} and imagery is given in Appendices~\ref{appendix:Asymmetric Galaxies}, \ref{appendix:Jellyfish Galaxies}, \ref{appendix:Truncated Galaxies}, and \ref{appendix:HDS-q}, to investigate the link between ongoing merger activity and RPS activity in A3667. While this sample size presents relatively low statistics for robust population-wide trends, it is notably larger than samples used in comparable single-cluster RPS studies (e.g., 5 galaxies in \citealt{Stroe2020} and \citealt{Foster2026}), and rivals the 45 candidates identified across the entire eight-cluster SAMI sample of \citet{Cakir2026}. This makes A3667 alone a sufficiently well-populated target for the analysis that follows, even within this early data release.

\begin{table}[!t]
    \centering
    \resizebox{0.75\linewidth}{!}{
    \begin{NiceTabular}{|l|c|}
        \Hline
        \textbf{Subsample} & $\bf N_{GAL}$ \\
        \Hline
        \Block{2-1}{Truncated galaxies (star-forming)} & \Block{2-1}{12} \\
         & \\
        \Hline
        Asymmetric galaxies (total) & 12 \\
         \textcolor{blue}{$\rightarrow$} HDSG (with central SF) & 5 \\
         \textcolor{blue}{$\rightarrow$} SFG & 6 \\
         \textcolor{blue}{$\rightarrow$} PASG & 1 \\
        \Hline
        Quenched HDSG (HDS-q) & 1 \\
        \Hline
        Jellyfish galaxies (total) & 3 \\
         \textcolor{blue}{$\rightarrow$} from \citet{Poggianti2016} & 2 \\
         \textcolor{blue}{$\rightarrow$} our visual inspection & 1 \\
        \Hline
    \end{NiceTabular}
    }
    \vskip 2mm
    \caption{The breakdown of the RPS sample used in this study.}
    \label{tab:RPS Sample}
\end{table}

\section{Analysis and Results}\label{sec:Analysis and Results}
\begin{figure*}[!ht]
    \centering
    \includegraphics[width=\textwidth]{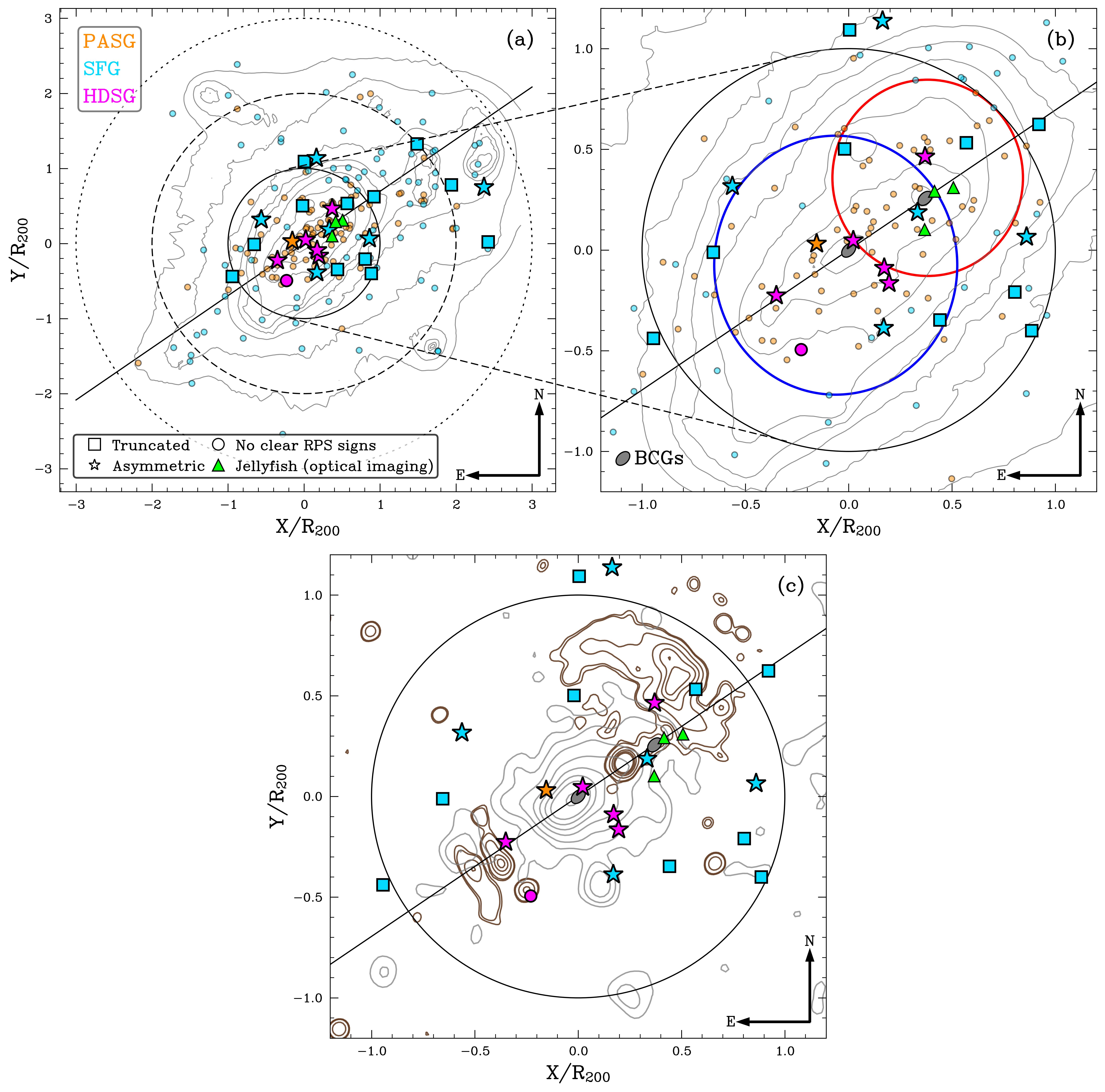}
    \caption{\textbf{(a)} Projected spatial distribution of A3667 cluster members with available Hector IFU and optical imaging data. Here, member positions are normalised by the cluster radius ($\rm R_{200}\simeq2.3$ Mpc; Owers et al, in preparation). Spectroscopic classifications of the members are represented by different colours: dark orange for PASGs, cyan for SFGs, and magenta for HDSGs. Symbols highlight different ionised gas morphologies: truncated galaxies---squares, asymmetric galaxies---stars, and circles for galaxies without clear RPS disturbance. The lime triangles represent the jellyfish candidates from the \citet{Poggianti2016} sample and the visual inspection outlined in Section~\ref{RPS candidates definition}. Grey contours illustrate the galaxy surface density determined through an adaptive kernel density estimation using the cluster members (Owers et al., in preparation), while the large solid, dashed, and dotted black circles mark $\rm R_{200}, \ 2R_{200}\text{, and } 3R_{200}$, respectively. \textbf{(b)} The same as (a), but zoomed in on the central $\rm R_{200}$. The brightest cluster galaxies of the main and the north-western clusters are indicated by grey-filled ellipses. The large blue and red circles mark the two primary substructures identified in \citet{Owers2009}: KMM5 (main cluster) and KMM2 (north-western cluster), respectively. \textbf{(c)} The same as (b), a close-up look at the central $\rm R_{200}$. Here, we only highlight the RPS candidates with the same colour code. The overlaid grey and dark brown contours represent X-ray emission from ROSAT, tracing the hot ICM, and radio emission from the GLEAM survey, which originates from accelerated relativistic electrons, respectively. The black solid line seen in all panels shows the approximate merger axis ($\theta_{x-axis} \sim\!35^\circ; \ \rm PA \sim\!-55^\circ $), defined as the line connecting the BCGs. For all panels, the North is oriented towards the top and East to the left, as shown by the arrows on the lower right.}
    \label{fig:Spatial distribution}
\end{figure*}
In Section~\ref{sec:Sample selection}, we selected a sample of galaxies exhibiting recent or current interaction with the ICM, likely through RPS. To understand if cluster merger activity enhances the rate of RPS activity, we now investigate the spatial and velocity distributions of RPS candidates with respect to substructures and merger features, such as shocks.

\subsection{Spatial Distribution}\label{subsec:Spatial Distribution}
As a first step, we examine the spatial distribution of RPS candidates relative to the general cluster population and merger features. Figure~\ref{fig:Spatial distribution} shows the spatial distribution of A3667 members. Figure~\ref{fig:Spatial distribution}(a) shows the projected spatial distribution of each population in A3667. Overall, the galaxy distribution is elongated along an SE--NW direction. The galaxy surface density contours reveal several substructures seen across the cluster, with the two main substructures---the main cluster (MC; large blue circle) and the NW subcluster (large red circle)---located at $\rm (X/R_{200}, Y/R_{200}) \simeq (0,0) $ and $\rm (X/R_{200}, Y/R_{200}) \simeq (0.37, 0.26)$, respectively. Considering the global spectral classes, it is clearly seen that PASGs (dark orange coloured symbols) are primarily concentrated towards the core of the system, while the cluster outskirts are dominated by SFGs. 

When focusing on RPS candidates (squares, stars, triangles, and the magenta point), which are shown in Figure~\ref{fig:Spatial distribution}(b) and (c), $\sim$\fraction{71}{10}{7}\footnote{Uncertainties on the fractions reported here and throughout the remainder of the paper are estimated following the recipe in \citet{Cameron2011}.} of the sample (20/28) are located within the central $\rm R_{200}$. Among them, fully/partially quenched asymmetric galaxies (PASG and HDSG) and optically selected jellyfish galaxies populate the inner cluster region, all located at $\rm R\lesssim0.6\ R_{200}$. These populations are primarily distributed along the merger axis connecting the BCGs ($\rm \theta_{x-axis} \sim\!35^\circ;\ position\ angle \sim\!-55^\circ$), situating them, in projection, entirely between the radio relics highlighted by the dark brown contours in Figure~\ref{fig:Spatial distribution}(c). In contrast, star-forming asymmetric and truncated galaxies show a wider spatial distribution and a milder spatial alignment with the merger axis. Most of these star-forming systems ($\sim$\fraction{90}{12}{4}; 16/18) are found at larger projected distances, $\rm R>0.5\ R_{200}$, with a median of $\sim\!0.92\ R_{200}$, compared to a median of $\sim\!0.4\ R_{200}$ for the quenched and jellyfish populations described above.

Furthermore, five asymmetric galaxies (including jellyfish galaxies; IDs = \texttt{C901005561300341, C901005481508430, C901005481506069, JO170, JO171}) appear concentrated in projection around the north-western BCG (BCG-NW; grey-filled ellipse within the large red circle). These systems are located downstream (south-east) of the outer NW radio relic (shock front). A second potential grouping of asymmetric galaxies is located near the MC core (around the main BCG; central grey-filled ellipse), a region that also hosts evidence of a turbulent ICM. We also identify four RPS candidates at the cluster outskirts ($\rm R > 2\ R_{200}$), with some being associated with infalling substructures. Hereafter, we refer to optically selected jellyfish galaxies/candidates also as ``asymmetric galaxies".

\subsection{Velocity Distribution}\label{subsec:Velocity Distribution}
In Section~\ref{subsec:Spatial Distribution}, we demonstrated a potential projected spatial alignment between the asymmetric galaxies, the cluster merger axis, and the merger-driven ICM features, which is particularly evident within the NW subsample. However, spatial distribution alone is insufficient to conclusively link these asymmetric galaxies to the ongoing merger activity. These positions can be supplemented by peculiar velocities to provide further kinematic context. Examining these velocities allows us to trace the relative motions of the cluster galaxies with respect to the driving forces of the merging substructures.
\\
\begin{figure}[t]
    \centering
    \includegraphics[width=\linewidth]{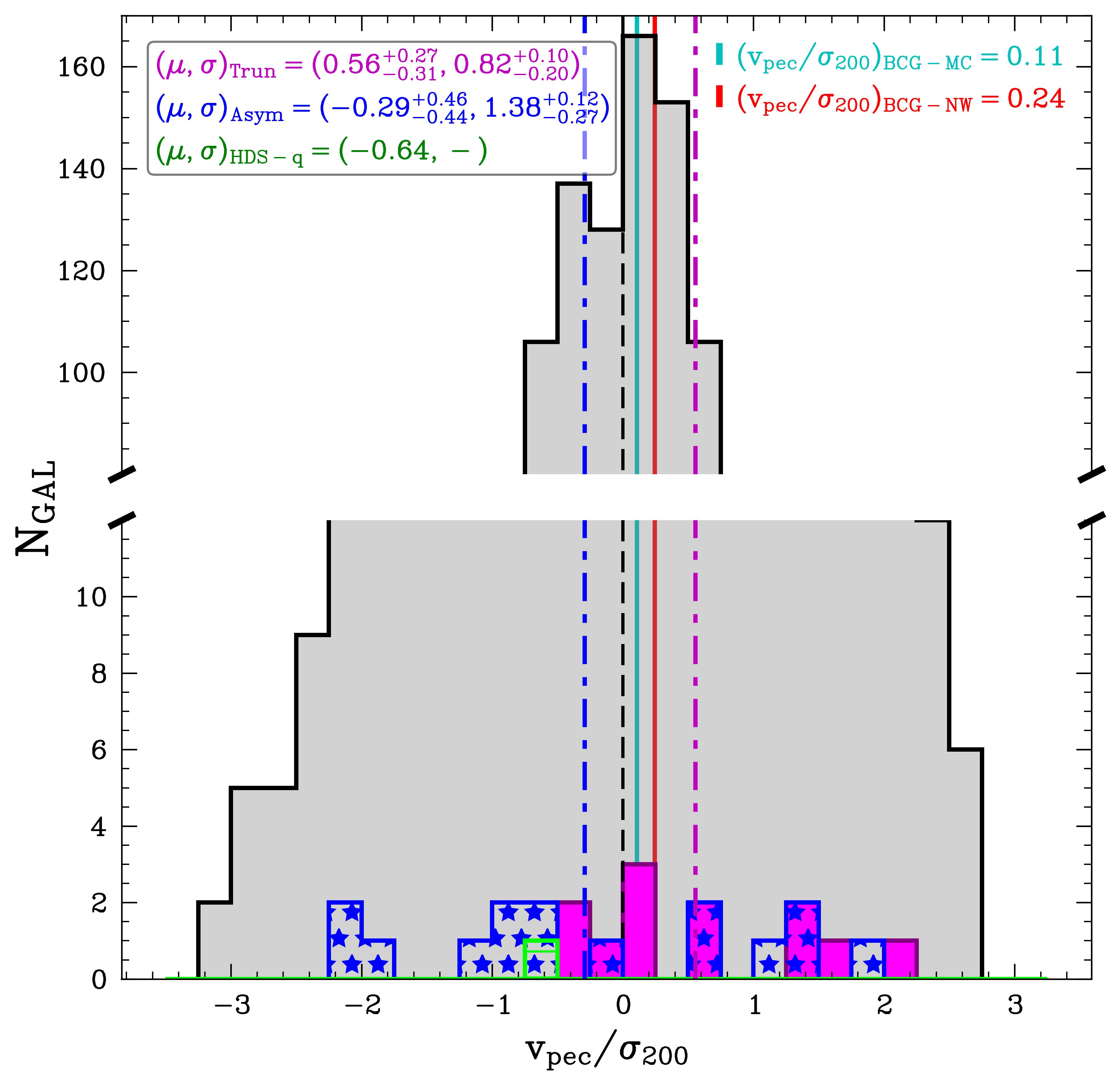}
    \caption{The distribution of normalised peculiar velocities ($\rm v_{pec}/\sigma_{200}$). The grey histogram shows the distribution for all members defined in Owers et al. (in preparation). The magenta, blue-starred, and green histograms represent the truncated, asymmetric, and HDS-q, respectively. In the upper left, we show the mean, $\mu$, and dispersion, $\sigma$, per sample determined using a biweight estimator \citep{Beers1990}. Uncertainties are estimated through a bootstrap approach, employing 1000 random resamplings ($\rm N_{iteration}=1000$). As there is only one HDS-q, there are no associated errors in the values. The black dashed vertical line is centred at 0, whereas the cyan and red vertical lines are centred at the velocities of the main BCG and the BCG-NW, respectively. The magenta and blue dashed vertical lines mark the biweighted means for the truncated and asymmetric sample, respectively. The RPS sample exhibits a wide distribution in normalised peculiar velocities relative to the cluster velocity dispersion ($\rm \sigma_{RPS} = 1.20^{+0.13}_{-0.16}\ \sigma_{200}$), with asymmetric galaxies showing the highest dispersion ($\rm \sigma_{Asym} = 1.38^{+0.12}_{-0.27}\ \sigma_{200}$), consistent with expectations for an infalling population.}
    \label{fig:Velocity histogram}
\end{figure}
\\
The distribution of normalised peculiar velocities, $\rm v_{pec}/\sigma_{200}$, is shown in Figure~\ref{fig:Velocity histogram}. We determine the mean ($\mu$) and dispersion ($\sigma$) for each subsample (values provided in the upper left of the figure), using biweight estimation \citep{Beers1990} with uncertainties derived from a bootstrap approach. The RPS sample exhibits a broader velocity distribution centred at $\rm \mu_{RPS} = 0.14^{+0.25}_{-0.27} \ \sigma_{200}$  compared to the overall cluster population, with a dispersion of $\rm \sigma_{RPS} = 1.20^{+0.13}_{-0.16}\ \sigma_{200}$. When broken down into subsamples, truncated galaxies notably appear to be redshifted with a mean velocity of $\rm \mu_{Trun} = 0.56^{+0.27}_{-0.31} \ \sigma_{200}$, while asymmetric galaxies are overall blueshifted ($\rm \mu_{Asym} = -0.29^{+0.46}_{-0.44} \ \sigma_{200}$). Among the subsamples, asymmetric galaxies exhibit the largest velocity dispersion,  $\rm \sigma_{Asym} = 1.38^{+0.12}_{-0.27}\ \sigma_{200}$. Conversely, truncated galaxies yield a narrower distribution ($\rm \sigma_{Trun} = 0.82^{+0.10}_{-0.20}\ \sigma_{200}$). Given the modest sample sizes of the truncated and asymmetric subsamples, these trends and their uncertainties should be interpreted with some caution. 

We compare the velocity distributions of RPS samples with the global A3667 population to test for statistically significant differences between the samples. For this purpose, we use the two-sample Anderson-Darling \citep{AD} and Kolmogorov-Smirnov \citep{KS} statistical tests, which evaluate the null hypothesis that these samples originate from the same parent distribution. Both tests yield p-values exceeding 0.05 ($\rm p \simeq 0.08 - 0.15$) for the truncated and asymmetric subsamples, indicating that they are likely drawn from the same parent velocity distribution. When we compare truncated and asymmetric samples, both KS and AD tests reveal a statistically significant difference ($\rm p_{KS} = 0.030,\ p_{AD}=0.047$). Given that the AD test is more sensitive to tails of a distribution, this significance is not likely driven by a few outliers.
\\
\begin{figure}[t]
    \centering
    \includegraphics[width=\linewidth]{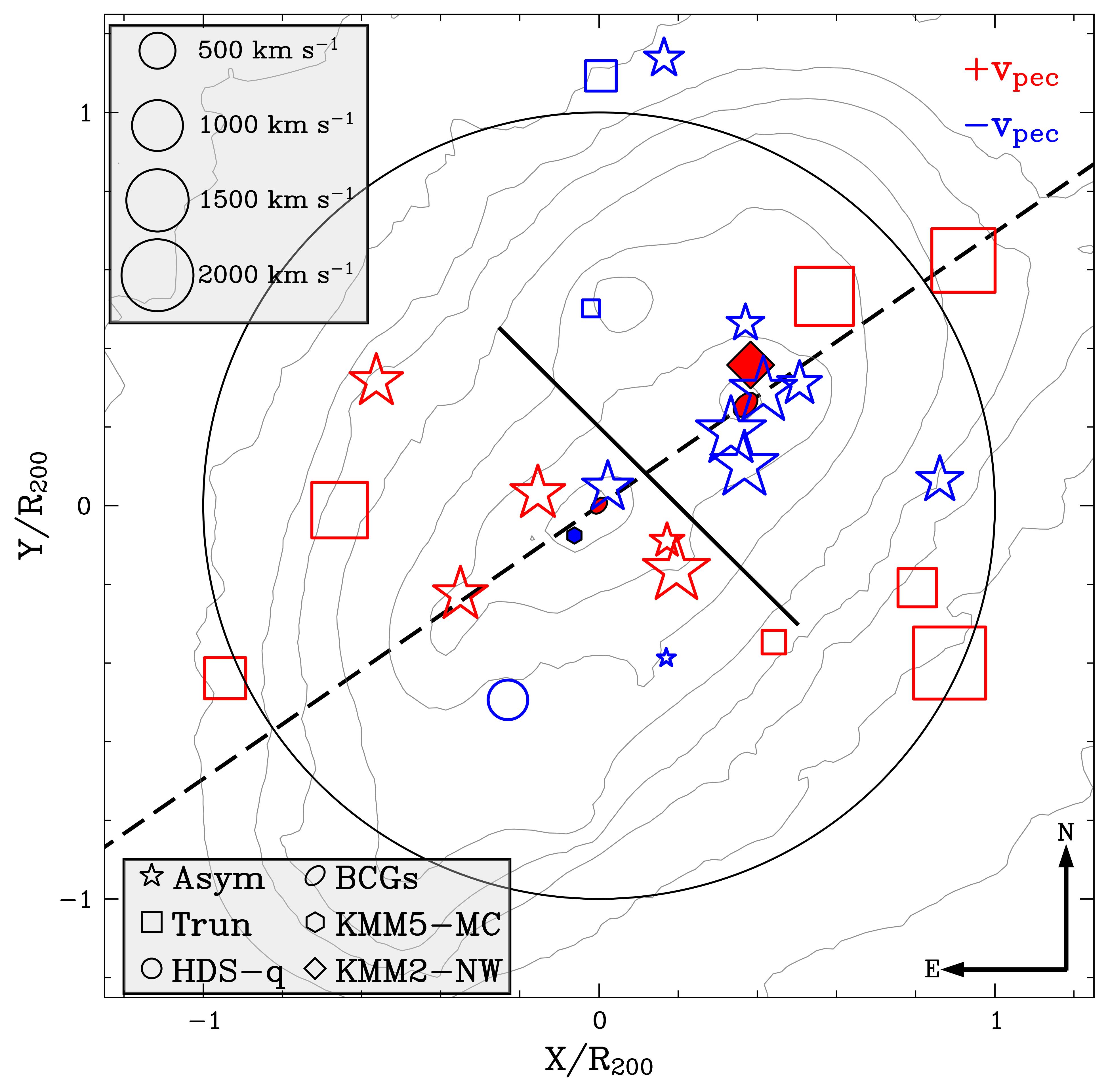}
    \caption{Spatial distribution of truncated (squares), asymmetric (including jellyfish galaxies; stars), and HDS-q (the circle) within $\rm \sim\!R_{200}$. The remaining symbols mark the location of substructures identified by \citet{Owers2009}: the BCGs (filled ellipses), the main cluster (KMM5-MC; filled hexagon) and the NW cluster (KMM2-NW; filled diamond). The colours represent the relative motion with respect to the cluster centre, adapted here; blue and red indicate negative and positive peculiar velocities, respectively. As shown in the upper left, the size of the symbol is proportional to the magnitude of the peculiar velocity. The dashed and solid black lines indicate the merger axis and the visually defined spatial boundary dividing A3667 into the main cluster (MC) and northwestern (NW) cluster regions, respectively. Asymmetric galaxies appear to be moving in the opposite direction with respect to the substructures with which they are spatially coincident in projection.}
    \label{fig:Spatial+velocity}
\end{figure}
\\
In Figure~\ref{fig:Spatial+velocity}, we combine the velocity information with the spatial distributions in order to examine the relative motions of RPS candidates with respect to substructures from \citet{Owers2009}, focusing on $\rm R \lesssim R_{200}$. The open squares and stars show the truncated and asymmetric galaxies, respectively, while the filled symbols mark the locations of the BCGs (pluses), the main cluster (KMM5-MC; hexagon), and the NW structure (KMM2-NW; diamond). The colours indicate the relative velocities of galaxies with respect to the cluster centre, with blue for $ v_{pec}<0 \ \rm km\ s^{-1}$ (i.e., blueshifted) and red for $v_{pec}>0 \ \rm  km\ s^{-1}$ (i.e., redshifted). The symbol sizes are proportional to the absolute value of the peculiar velocities. The dashed black line indicates the merger axis, while the solid black line represents a visual boundary used to divide A3667 into the ``MC" and ``NW" regions, focusing particularly on the central $\rm R_{200}$ region for subsequent analysis.

Considering the asymmetric galaxies, when combined with their spatial distribution, they present two spatially localised velocity structures. All NW asymmetric galaxies are blueshifted relative to the cluster centre (i.e., $v_{pec}\lesssim-580 \ \rm km\ s^{-1}$), with median $v_{pec}\sim\!-1408 \ \rm km\ s^{-1}$. In contrast, the median peculiar velocity of MC counterparts is $\rm \sim\!1182 \ \rm km\ s^{-1}$ (i.e., redshifted). Five out of seven MC asymmetric galaxies have $v_{pec}\gtrsim-140 \ \rm km\ s^{-1}$, indicating that the NW and MC galaxies are separated in the velocity space. Furthermore, the comparison between the peculiar velocities of the main subclusters adopted from \citet{Owers2009} (i.e., $ v\rm_{KMM5,\ MC} = -103 \  km\ s^{-1}$ and $v\rm_{KMM2,\ NW} = 422 \ km\ s^{-1}$) and the median peculiar velocities of NW and MC asymmetric galaxies suggests that the asymmetric galaxies move in the opposite direction with respect to the subclusters with which they coincide in projection, indicating that they are not associated with the merging substructures.

\subsection{Tail orientations}\label{subsec:Tail Orientations}
Section~\ref{subsec:Velocity Distribution} showed that asymmetric galaxies do not seem to co-move with the substructures with which they are spatially coincident in projection. To investigate the idea that they originated from an infalling population, we lastly examine the tail orientations.
\\
\\
As shown in Appendix~\ref{appendix:Formulas}, we define three different tail orientations, $\rm \theta_{tail,c}, \theta_{tail,m}\text{, and }\theta_{tail,W}$, as follows
\begin{itemize}
    \item \textcolor{red}{$\bf \theta_{tail,c}$} is the angle between \textit{i)} the tail vector ($\vec{\rm T}$) connecting the galaxy's ionised gas centre to the galaxy's stellar centre, and \textit{ii)} the cluster-centric vector ($\vec{\rm R}$) connecting the galaxy's stellar centre to the cluster centre. In this convention, $\rm \theta_{tail,c}=180^{\circ}$ indicates a tail pointing directly away from the cluster centre, while $\rm \theta_{tail,c}=0^{\circ}$ indicates a tail pointing directly towards the centre.
    \vskip 2mm
    \item \textcolor{red}{$\bf \theta_{tail,m}$} is defined as the angle between \textit{i)} the tail vector and  \textit{ii)} the merger axis vector ($\vec{\rm M}$) connecting the BCGs' stellar centres. Here, we consider the absolute misalignments, in which $\rm \theta_{tail,m}=0^{\circ}$ means the tail is perfectly aligned with the axis, whereas $\rm \theta_{tail,m}=90^{\circ}$ indicates the tail is perpendicular to the axis.
    \vskip 2mm
    \item \textcolor{red}{$\bf \theta_{tail, W}$} is defined as the angle between the tail vector and the x-axis. Under this convention, angles of $0^\circ, +90^\circ, \pm180^\circ$, and $-90^\circ$ correspond to the West (W), North (N), East (E), and South (S) cardinal directions, respectively. 

\end{itemize}
\noindent Tail orientations for the jellyfish galaxies from the \citet{Poggianti2016} sample are adopted from \citet{Salinas2024}, including the counter-clockwise angle from the x-axis (West)---x-axis angle---and the positional angle relative to the BCG (cluster centre). We calculate the orientation relative to the merger axis by subtracting the merger axis angle from the tail's x-axis angle. For the jellyfish galaxy identified via Legacy imaging, we adopt a similar method to the Hector sample. As shown in Appendix \ref{fig: LS Jellyfish}, we isolate the galaxy in the $g$-band image and apply a manual threshold in flux to include tails. We subsequently employ a centroid-based proxy to approximate both the gas centre and the resulting tail orientation. Notably, as tail orientations are measured in 2D, our interpretations are inherently subject to projection effects for the remainder of the paper.
\\
\begin{figure*}[!ht]
    \centering
    \includegraphics[width=\linewidth]{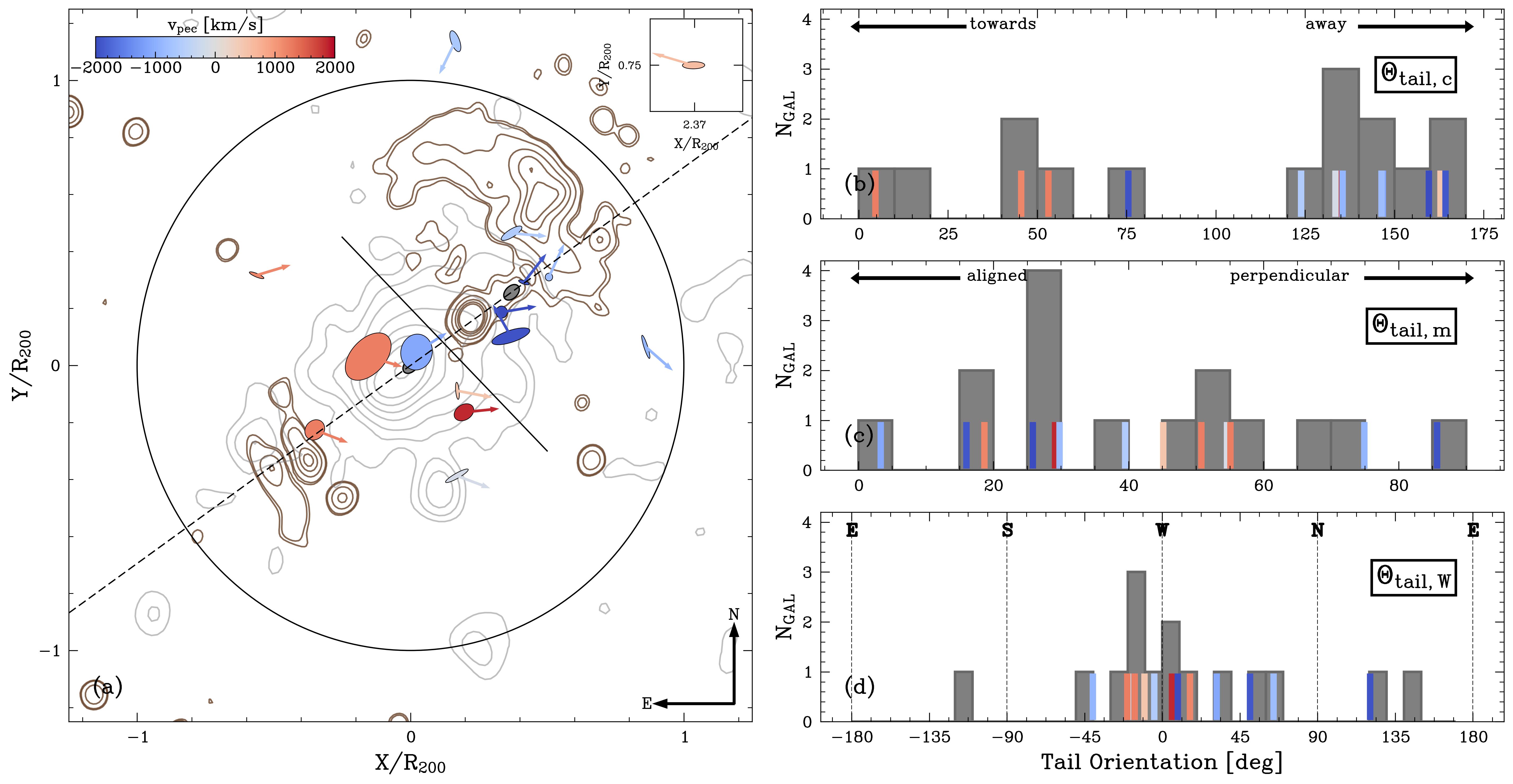}
    \caption{\textbf{Left panel:} Spatial distribution of the asymmetric galaxies (ellipses) with a close-up view of the $\rm R_{200}$ (black circle), overlaid on X-ray (grey) and radio (dark brown) emission contours. Arrows indicate the orientations of the ionised gas tails, connecting the stellar and gas centres. Ellipse sizes are proportional to the effective radii (magnified by a factor of 15), while arrow lengths are fixed to an arbitrary scale. Ellipses and arrows are colour-coded by peculiar velocity. The dashed and solid black lines represent the approximate merger axis and demarcation boundary between the NW and MC cluster regions, respectively. \textbf{Right panel:} Distribution of tail orientations (grey histogram) with respect to the cluster centre ($\theta_{\rm tail,c}$; \textbf{top}), the merger axis ($\theta_{\rm tail,m}$; \textbf{middle}), and the West ($\theta_{\rm tail,W}$; \textbf{bottom}). Individual values for asymmetric galaxies within $\rm R_{200}$ are marked by stripes, following the colour scheme of the left panel. The letters S, W, N, and E denote the cardinal directions.}
    \label{fig:Tail orientations}
\end{figure*}
\\
Figure~\ref{fig:Tail orientations}(a) presents the same asymmetric galaxies as in Figure~\ref{fig:Spatial distribution}, now with their ionised gas tails marked, defined as the vector connecting each galaxy's centre to its emission centre. The histograms in the right column show the distribution of tail orientations relative to the cluster centre (top panel), the merger axis (middle panel), and the West (i.e., $+\rm x-axis$; bottom panel) for all asymmetric galaxies. The stripes represent the individual values for asymmetric galaxies within $\rm R_{200}$ and are colour-coded to their peculiar velocities. To quantify tail orientations relative to the cluster centre, we adopt the geometric framework of \citet{Salinas2024}, defining tails as pointing ``towards" ($\theta_{\rm tail,c} < 45^{\circ}$), ``perpendicular" to ($45^{\circ} \le \theta_{\rm tail,c} \le 135^{\circ}$), or ``away" from ($\theta_{\rm tail,c} > 135^{\circ}$) the cluster centre. As shown in Figure~\ref{fig:Tail orientations}(b), the full asymmetric sample ($\rm N=15$) yields an overall median orientation of $\sim\!135^{\circ}$ with an angular dispersion of $\sigma_{\rm MAD}(\theta_{\rm tail,c}) \sim\!35^{\circ}$ (where $\sigma_{\rm MAD} = 1.4826 \times \text{median absolute deviation, MAD}$). This complete population shows a mixed distribution across the away ($\rm N=6$; $\sim$\fraction{40}{11}{13}) and perpendicular ($\rm N=7$; $\sim$\fraction{47}{12}{12}) regimes, with only two galaxies pointing towards the core (2/15; $\sim$\fraction{13}{5}{13}).

When considering the central $\rm R_{200}$, the NW subsample ($\rm N=6$) yields a median orientation of $\sim141^{\circ}$ with a narrower angular dispersion of $\sigma_{\rm MAD} \sim\!27^{\circ}$, and it is strictly split between away ($\rm N=4$; $\sim$\fraction{67}{21}{13}) and perpendicular ($\rm N=2$; $\sim$\fraction{33}{13}{21}) tails, without any tails pointing towards the cluster centre. Conversely, the MC sample ($\rm N=7$) exhibits a median orientation of $\sim\!133^{\circ}$ but with a larger angular dispersion ($\sigma_{\rm MAD} \sim\!44^{\circ}$). This MC population is predominantly characterised by perpendicular alignments ($\rm N=4$; $\sim$\fraction{57}{18}{15}), with two galaxies (2/7; $\sim$\fraction{29}{11}{20}) oriented radially away and one oriented towards the core (1/7; $\sim$\fraction{14}{5}{21}).

Figure~\ref{fig:Tail orientations}(c) shows that $\sim$\fraction{54}{13}{12} (7/13) of asymmetric tails within $\rm R_{200}$ have $\theta_{\rm tail,m}>30^\circ$, indicating no strong population-wide preference for alignment with the merger axis. This threshold, while heuristic, is commonly used in gas misalignment studies \citep[e.g.,][]{Bryant2019, Ristea2022}. However, a clearer trend emerges when considering the two subregions separately. In the MC region, the median tail--merger axis angle is $\sim\!45^\circ$ with a relatively low dispersion ($\rm \sigma_{MAD}\simeq15^\circ$), suggesting the MC galaxies are broadly misaligned with the merger axis. In contrast, $\sim$\fraction{67}{21}{13} (4/6) of galaxies in the NW subsample exhibit mild to strong alignment with the merger axis ($\rm \theta_{tail,m}<40^\circ$), with a lower median of $\sim\!35^\circ$ and a slightly larger dispersion ($\rm \sigma_{MAD}\simeq20^\circ$) mainly driven by 2 galaxies. Notably, these four NW galaxies are situated in close projection to the NW radio relic, suggesting a potential connection to merger-driven activity in that region.

The tail directions displayed in Figure~\ref{fig:Tail orientations}(a) and quantified in Figure~\ref{fig:Tail orientations}(d) indicate that 10 out of 13 tails ($\sim$\fraction{77}{15}{8}) are within $\pm45^\circ$ of due west, with a median of $\rm (\theta_{tail,W})\simeq6^\circ$. This trend is particularly clear for the MC sample, with $\rm (\theta_{tail,W})_{MC,median}\simeq-10^\circ$ and $\rm \sigma_{MAD}(\theta_{tail,W})_{MC}=15^\circ$. In contrast, the NW sample shows a shift away from due west towards the northwest, with $\rm (\theta_{tail,W})_{NW,median}\simeq30^\circ$ and $\rm \sigma_{MAD}(\theta_{tail,W})_{NW}\simeq51^\circ$, alongside a much larger spread in orientation.

Lastly, two-sample AD and KS tests reveal no statistically significant difference between the MC and NW distributions for all of the above orientation measures ($\rm p\gg0.05$), indicating that, despite the differences in median and scatter described above, the overall shapes of the distributions are not formally distinguishable given the current sample sizes.

\section{Discussion}\label{sec:Discussion}
The primary aim of this study is to investigate the potential for merger-induced activity in galaxy properties through RPS. On a large scale, the spatial distribution shown in Section~\ref{subsec:Spatial Distribution} and Figure \ref{fig:Spatial distribution} reveals that the majority of RPS candidates (20/28) are found within $\rm R_{200}$, in agreement with previous studies \citep{Smith2010, Yoon2017, Jaffe2018, Jung2018, Owers2019, Roman-Oliveira2019, RP2020, Stroe2020, Roberts2021b, Roberts2021a, Bellhouse2022,  Kolcu2022, Salinas2024, Poggianti2025, Cakir2026}, which found that galaxies with asymmetric tails, truncated disks, or strong Balmer absorption, indicative of recently quenched star formation, are preferentially located close to the cluster centre. Combined with this spatial configuration, the asymmetric galaxies also exhibit a broader peculiar velocity distribution than the overall cluster population, with a dispersion of $\sigma_{\rm Asym} \simeq 1.4\sigma_{200}$, consistent with a population of recent infallers \citep{Rhee2017, Jaffe2018, Owers2019, Cakir2026}.

The RPS candidates identified in the outskirts ($\rm R > R_{200}$) are predominantly truncated galaxies (\fraction{75}{19}{9}; 6/8). Of this outskirts population, four RPS candidates (three truncated and one asymmetric) are located at $\rm R>2\ R_{200}$, of which three are likely associated with nearby substructures, supporting the pre-processing scenario \citep{Fujita2004, McGee2009, Mahajan2012, Haines2015, Rhee2020, Piraino-Cerda2024}. RPS is active not only at the cluster cores; it can also operate efficiently at galaxy group densities \citep{Westmeier2017, Brown2017, Kolcu2022}. Therefore, the detection of galaxies with disturbed gas morphologies at such large radii demonstrates they are undergoing active group-scale stripping, which pre-processes these systems before they ever accrete onto the main cluster. For example, \citet{Piraino-Cerda2024} investigated RPS activity in an interacting cluster system, A2670, where they identified 101 candidates disturbed by RPS out to $\rm 5\ R_{200}$. The RPS candidates found at larger cluster-centric radii were mainly associated with the substructures, and the denser substructures exhibited more disturbed systems. Our findings align with these previous studies and indicate that actively assembling cluster regions can affect galaxies out to larger cluster-centric distances.
\\
\\
Truncated galaxies are often interpreted as a post-RPS population that has likely survived a core passage \citep{Koopmann2004, Chung2009, Yoon2017, Owers2019, Cakir2026}. As shown in Figure~\ref{fig:Spatial distribution}, truncated galaxies are absent from the central region of A3667 and instead reside at larger cluster-centric distances 
($\rm R/R_{200} \gtrsim 0.5$), suggesting they represent a more evolved stage of RPS compared to the asymmetric population. This is supported by \citet{Cakir2026}, who found that the majority of SAMI cluster galaxies with truncated ionised gas disks (\fraction{63}{9}{8}; 19/30) are located at $\rm R/R_{200}\gtrsim0.4$. We also detect a deficit of truncated galaxies between $\rm 1-2\ R_{200}$, where backsplash galaxies are expected \citep{Gill2005}, in contrast to the truncated \textsc{Hi} disks identified at similar radii in Virgo by \citet{Yoon2017}. This deficit may partly reflect incomplete spatial coverage of the A3667 outskirts in the current data release, but may also be a consequence of the dynamically young environment of A3667, where unrelaxed systems are expected to host fewer backsplash galaxies than relaxed counterparts \citep{Haggar2020}. Alternatively, the truncated galaxies may represent a descendant population of the MC asymmetric galaxies, having undergone a more advanced stage of RPS following an earlier accretion event, which would naturally account for their spatial distribution at larger cluster-centric radii (Section~\ref{subsec:Spatial Distribution}) and high positive peculiar velocities (Section~\ref{subsec:Velocity Distribution}).
\\
\\
Zooming in on the central cluster region in Figure~\ref{fig:Spatial distribution}(c), the asymmetric galaxies, in particular those with extreme stripping signatures, exhibit a narrow spatial distribution elongated along the merger axis and are primarily located between the NW and SE radio relics (dark brown contours). A similar spatial alignment has also been reported by \citet{Stroe2020} for galaxies with gas disturbance in a post-merger system at $z\sim0.2$, in line with our findings and raising a potential link to merger-driven activity. The presence of a clump of asymmetric galaxies found close in projection to the NW substructure and downstream of the NW radio relic strengthens the possibility of merger-enhanced RPS activity. Moreover, peculiar velocities presented in Figure~\ref{fig:Spatial+velocity} indicate that NW asymmetric galaxies ($\rm median(v_{pec}) \simeq -1408 \ km\ s^{-1}$) are not associated with the NW substructures ($\rm median(v_{pec})= 422 \ \rm km\ s^{-1}$). Given the larger difference in velocities ($\rm |\Delta v_{pec}| \sim\!2000 \ \rm km \ s^{-1}$), the extreme RPS signatures observed in these galaxies are likely driven by the high-velocity encounter with the ICM of the NW substructure, aligning with the findings reported by \citet{Owen2006}, \citet{Owers2012}, \citet{Ebeling2019}, and \citet{Bellhouse2022} for other merging systems.
\\
\\
The gas tails investigated in Section~\ref{subsec:Tail Orientations} show a tentative preference for pointing away from the cluster centre in the NW subcluster region, which is consistent with a pre-pericentre, infalling population \citep{Chung2007, Smith2010, RP2020, Roberts2021a, Kolcu2022, Smith2022, Salinas2024}. However, this trend is not clearly present across the full sample or the MC region, where the distribution of tail orientations is more ambiguous, an ambiguity compounded by the 2D projection effects noted in Section~\ref{subsec:Tail Orientations}. Our results for the full sample are broadly consistent with \citet{Roman-Oliveira2019, Roman-Oliveira2021} and \citet[as in their Figure~A2]{Salinas2024}, who similarly found no preferential pattern of tail orientation in merging systems: the former in the pre-merger multi-cluster A901/2 system, where four substructures are in coalescence, and the latter across nine interacting clusters spanning a range of assembly stages. Using tailored simulations of the same A901/2 system, \citet{Ruggiero2019} found that jellyfish galaxies tend to emerge at the leading edges of haloes owing to ICM pressure jumps driven by halo--halo interactions. This absence of a coherent trend therefore reflects the diverse merger geometries inherent to such systems, which can dilute any detectable orientation signal when subclusters or clusters are stacked together, e.g., as is also seen in the relaxed sample of \citet{Salinas2024}.

To search for direct evidence of merger-elevated RPS activity, we also examined the misalignments between gas tails and the merger axis (Section~\ref{subsec:Tail Orientations}). In the NW region, $\sim\!67\%$ (4/6) of tails exhibit mild to strong alignment with the merger axis ($\rm \theta_{tail,m}<40^\circ$), consistent with the tail--merger axis alignments reported in previous studies \citep{Rawle2014, Stroe2020, Edler2026} and supporting merger-driven RPS activity in that region. In contrast, the MC region shows a higher median misalignment ($\rm \theta_{tail,m}\sim45^\circ$), which does not follow the same trend. This discrepancy may reflect the unique dynamical configuration of the merger, where the MC and NW substructures are at different stages of interaction, potentially giving rise to distinct local ICM conditions and ram-pressure geometries in each subregion.

Beyond the cluster centre and merger axis orientations, Figure~\ref{fig:Tail orientations}(d) reveals that tails preferentially point towards the West in the MC region and towards the northwest in the NW subcluster region, suggesting a coherent motion originating from a shared direction, likely a nearby filament accreting. Whilst this is not consistent with an isotropic infall scenario \citep{Lotz2019, RP2020, Roberts2021b}, similar directional coherence has been reported by \citet{Ebeling2019} for the post-merger cluster A1758N ($z\sim0.28$), where six out of eight jellyfish galaxy tails point preferentially towards the SE. Altogether, combining this directional bias with negative radial velocities (blueshifted) of these jellyfish galaxies, \citet{Ebeling2019} suggest a bulk infall along an attached filament, rather than isotropic infall. Their results show similarities with our findings in Section~\ref{subsec:Tail Orientations} and Figure~\ref{fig:Tail orientations} for the tail directions of our asymmetric galaxy sample, supporting a filamentary accretion scenario instead of isotropic infall.
\\
\\
In light of our findings and comparisons with the literature, we conclude that the asymmetric galaxies in A3667 were most likely accreted onto the cluster from a nearby filament along the NW-W direction. Given the extreme stripping signatures, the NW sample currently experiences elevated RPS activity by interacting with the merger-driven high-velocity ICM environment. \citet{Moretti2018} also argued this merger-driven RPS scenario for JO171 because of its proximity to the radio bridge \citep{Carretti2013} in the NW cluster core. Regarding the filamentary accretion scenario, direct evidence indeed comes from recent eROSITA X-ray observations. \citet{Dietl2024} recently discovered a $\sim$13 Mpc long, in projection, X-ray filament connecting A3667 and another nearby cluster A3651 ($z\simeq0.06$), and it is connected to A3667 from the NW direction (Figure 2 of \citealt{Dietl2024}). Redshifts of galaxies in the filament show an increasing progression from A3667 to A3651 (Figure 7 of \citealt{Dietl2024}), with galaxies spatially close to each cluster having similar redshifts, likely indicating motion towards the clusters. Given that A3651 has a higher cosmological redshift and is therefore more distant than A3667, the filament is probably connected to A3667 from behind, with A3651 located further along the line of sight. This inferred 3D geometry, an accretion filament approaching A3667 from behind and matching the blueshifted NW velocities, provides direct evidence against isotropic infall for the NW population, complementing our tail orientation results with an independent, kinematically motivated view of the accretion geometry.

However, the MC asymmetric galaxies are difficult to explain within this NW-filament-merger interpretation. The preference for tails to be oriented to the west and the negative peculiar velocities of the MC asymmetric galaxies indicate that their true 3D motion is distinct from the NW asymmetric galaxies. Moreover, the properties of the MC asymmetric galaxies appear to also be distinct from their NW counterparts. In particular, the spectroscopic classification and ionised gas maps of the MC galaxies shown in Appendices~\ref{appendix:Asymmetric Galaxies} and \ref{appendix:Jellyfish Galaxies} reveal severe truncation in their gas disks and strong H$\delta-$strong regions, pointing to a relatively advanced stage of RPS. In contrast, NW galaxies are primarily jellyfish galaxies with star-forming tails, indicative of earlier stages of ongoing RPS, where the majority of the gas reservoir is still retained. The main cluster also shows evidence for bulk motions of the ICM in the form of cold fronts \citep{Vikhlinin2001, Owers2009} and plumes of ICM \citep{Mazzotta2002}, which lends support for merger-related amplification of RPS in the core of the cluster. These pieces of evidence suggest that the MC asymmetric galaxies may have been accreted from a westerly direction in a separate accretion event from the NW asymmetric galaxies, into the dynamically turbulent ICM environment found in the MC region.
\\
\\
Our results suggest a scenario of merger-enhanced RPS activity within the core of A3667, primarily affecting a galaxy population accreted from nearby filaments. Considering that this elevated activity is not associated with the galaxies of merging substructures themselves, our conclusion is in line with previous studies \citep{Cohen2014, Cohen2015, Cakir2025} which suggest that elevated star-forming galaxy fractions result from the spatial mixing of distinct galaxy populations, such as infalling group members and filament galaxies, accreted into dynamically active cluster environments. Consequently, this study provides direct evidence for elevated RPS activity in the merger-affected regions, offering a physical explanation for these broader observational trends.

\section{Summary and Conclusions}\label{sec:Conclusion}
Whether cluster mergers accelerate the evolution of galaxies through hosting more hostile environments or not remains an open question. In this paper, we address this question by identifying galaxies exhibiting environment-driven quenching signatures in a nearby merging cluster, A3667, primarily through ram pressure stripping. Utilising the spatially resolved spectroscopy from the Hector Galaxy Survey, we defined a sample of RPS candidates based on truncations and asymmetries in the ionised gas distributions---truncated and asymmetric galaxies. Also, we expanded the asymmetric galaxy sample by combining it with the jellyfish galaxies identified by previous studies and one newly identified through visual inspection of Legacy Survey imaging. We subsequently analysed the spatial distributions (Section~\ref{subsec:Spatial Distribution}), the peculiar velocities (Section~\ref{subsec:Velocity Distribution}), and tail orientations (Section~\ref{subsec:Tail Orientations}) of the RPS candidates. The key findings are the following:

\begin{enumerate}
    \item We identified 28 RPS candidates in Abell 3667, extending out to $\rm 3\ R_{200}$. This sample significantly increases the number of RPS-affected galaxies in this cluster compared with the previously known RPS population from large observational campaigns.
    \vskip 2mm 
    \item The majority of RPS candidates ($\sim$\fraction{71}{10}{7}; 20/28) are found within $\rm R_{200}$, with asymmetric galaxies exhibiting extreme stripping predominantly distributed along the merger axis, between the NW and SE radio relics. Among these asymmetric galaxies, we identify two sub-groups, located at the NW substructure and the main cluster (MC) in projection, both showing evidence of turbulent environments with high-velocity bulk ICM motions, supporting merger-driven amplification of RPS activity in these regions.
    \vskip 2mm    
    \item The asymmetric sample presents a wider distribution of the peculiar velocity relative to the cluster population (i.e., $ \sigma_{Asym} = 1.38^{+0.12}_{-0.27}\ \sigma_{200}$), consistent with an infalling population, whereas truncated galaxies are predominantly redshifted with smaller velocity dispersion  ($0.82^{+0.10}_{-0.20}\ \sigma_{200}$). When spatially analysed, asymmetric galaxies exhibit two spatially clustered velocity structures: the blueshifted NW and redshifted MC asymmetric galaxies. Both subsamples appear not to co-move with the substructures in their vicinity, indicating that they are likely not members of either substructure.    
    \vskip 2mm 
    \item The tails of asymmetric galaxies show a mild bias towards pointing away from the cluster centre ($\rm median(\theta_{tail,c}) \sim\!133^\circ$), though this trend is clearer in the NW subcluster region than in the MC region. When comparing tail orientation to the merger axis, the NW region shows a preference for mild to strong alignment ($\sim\!67\%$, 4/6; $\rm \theta_{tail,m}<40^\circ$), indicating merger-driven stripping activity, whereas the MC region shows a higher median misalignment ($\rm \theta_{tail,m}\sim\!45^\circ$). Furthermore, the tails predominantly point towards the west ($\sim\!77\%$, 10/13), with the MC tails closely aligned with due west and the NW tails shifted slightly towards the northwest, likely tracing a bulk infall from a nearby filament along the NW-W direction.
\end{enumerate}

\noindent These results provide evidence for merger-enhanced RPS activity in the centre of A3667, arising from multiple infalling populations caught up in an ongoing merger. The NW asymmetric sample, which has emerged in the form of jellyfish morphologies, is currently undergoing extreme RPS, likely driven by the large-scale bulk ICM motion associated with the NW merging subcluster. The MC asymmetric galaxies, on the other hand, appear to have arrived earlier than their NW counterparts and are at a more advanced stage of RPS, with their distinct kinematics and the surrounding turbulent ICM environment pointing to a separate, merger-related accretion event from the west.
\\
\\
To examine the impact of the merger in greater detail, the star formation activity of these galaxies needs to be investigated through both integrated and spatially resolved star formation profiles, as well as star formation histories, in order to quantify any merger-induced star formation activity and to determine quenching timescales relative to the merger timescale. In this early science paper, we demonstrated the capability of the Hector instrument to address such questions. The ongoing Hector Galaxy Survey \citep{Bryant2024, Oh2025} will deliver data for 11 massive clusters spanning a range of merger activity (including A3667), enabling a direct comparison between merging and relaxed clusters in future work.
\\
\\
\small \noindent\textbf{\textcolor{largeblue}{Acknowledgment.}} OÇ and GQ acknowledge the support from the Commonwealth through an Australian Government Research Training Program Scholarship (\textcolor{largeblue}{https://doi.org/10.82133/C42F-K220}). This research was partially supported by the Australian Research Council Centre of Excellence for All Sky Astrophysics in 3 Dimensions (ASTRO 3D), through project number CE170100013. This work was supported by the Korea Astronomy and Space Science Institute under the R\&D program (Project No. 2025-1-831-01) supervised by the Ministry of Science and ICT (MSIT). MP and JHL acknowledge support from the National Research Foundation of Korea (NRF) grant funded by the Korea government (MSIT) (No. 2022R1A2C1004025). CF is the recipient of an Australian Research Council Future Fellowship (project number FT210100168) funded by the Australian Government. CF is a recipient of ARC Discovery Project DP210101945. SO acknowledges support from the Korean NRF (RS-2023-00214057; RS-2025-00514475). KO acknowledges support from the Korea Astronomy and Space Science Institute under the R\&D program (Project No. 2026-1-831-01), supervised by the Korea AeroSpace Administration, and the National Research Foundation of Korea (NRF) grant funded by the Korea government (MSIT) (RS-2025-00553982). 

The Hector Galaxy Survey is based on observations made at the Anglo-Australian Telescope. We acknowledge the traditional owners of the land on which the AAT stands, the Gamilaraay people, and pay our respects to elders past and present. The Hector multi-object integral field spectrograph instrument was built jointly by the University of Sydney and Macquarie University nodes of the Astralis Astronomical Instrumentation Consortium (\textcolor{largeblue}{https://astralis.org.au/}), with additional financial contributions from the Australian National University and University of Western Australia and supported by the Australian Research Council through grants LE170100242, LE190100018 and FT180100231. The Hector input catalogue is based on data taken from the WAVES Survey, Sloan Digital Sky Survey, GAMA Survey, 2dfGRS and Skymapper Southern Sky Survey. The Hector Galaxy Survey research was supported by the Australian Research Council Centre of Excellence for All Sky Astrophysics in 3 Dimensions (ASTRO3D), through project number CE170100013, and other participating institutions. The Hector Galaxy Survey website is \textcolor{largeblue}{https://hector.survey.org.au/}. The Hector Galaxy Survey makes use of Data Central services (\textcolor{largeblue}{datacentral.org.au}).

The Legacy Surveys consist of three individual and complementary projects: the Dark Energy Camera Legacy Survey (DECaLS; Proposal ID \#2014B-0404; PIs: David Schlegel and Arjun Dey), the Beijing-Arizona Sky Survey (BASS; NOAO Prop. ID \#2015A-0801; PIs: Zhou Xu and Xiaohui Fan), and the Mayall z-band Legacy Survey (MzLS; Prop. ID \#2016A-0453; PI: Arjun Dey). DECaLS, BASS and MzLS together include data obtained, respectively, at the Blanco telescope, Cerro Tololo Inter-American Observatory, NSF’s NOIRLab; the Bok telescope, Steward Observatory, University of Arizona; and the Mayall telescope, Kitt Peak National Observatory, NOIRLab. Pipeline processing and analyses of the data were supported by NOIRLab and the Lawrence Berkeley National Laboratory (LBNL). The Legacy Surveys project is honoured to be permitted to conduct astronomical research on Iolkam Du’ag (Kitt Peak), a mountain with particular significance to the Tohono O’odham Nation.
NOIRLab is operated by the Association of Universities for Research in Astronomy (AURA) under a cooperative agreement with the National Science Foundation. LBNL is managed by the Regents of the University of California under contract to the U.S. Department of Energy.
This project used data obtained with the Dark Energy Camera (DECam), which was constructed by the Dark Energy Survey (DES) collaboration. Funding for the DES Projects has been provided by the U.S. Department of Energy, the U.S. National Science Foundation, the Ministry of Science and Education of Spain, the Science and Technology Facilities Council of the United Kingdom, the Higher Education Funding Council for England, the National Center for Supercomputing Applications at the University of Illinois at Urbana-Champaign, the Kavli Institute of Cosmological Physics at the University of Chicago, Center for Cosmology and Astro-Particle Physics at the Ohio State University, the Mitchell Institute for Fundamental Physics and Astronomy at Texas A\&M University, Financiadora de Estudos e Projetos, Fundacao Carlos Chagas Filho de Amparo, Financiadora de Estudos e Projetos, Fundacao Carlos Chagas Filho de Amparo a Pesquisa do Estado do Rio de Janeiro, Conselho Nacional de Desenvolvimento Cientifico e Tecnologico and the Ministerio da Ciencia, Tecnologia e Inovacao, the Deutsche Forschungsgemeinschaft and the Collaborating Institutions in the Dark Energy Survey. The Collaborating Institutions are Argonne National Laboratory, the University of California at Santa Cruz, the University of Cambridge, Centro de Investigaciones Energeticas, Medioambientales y Tecnologicas-Madrid, the University of Chicago, University College London, the DES-Brazil Consortium, the University of Edinburgh, the Eidgenossische Technische Hochschule (ETH) Zurich, Fermi National Accelerator Laboratory, the University of Illinois at Urbana-Champaign, the Institut de Ciencies de l’Espai (IEEC/CSIC), the Institut de Fisica d’Altes Energies, Lawrence Berkeley National Laboratory, the Ludwig Maximilians Universitat Munchen and the associated Excellence Cluster Universe, the University of Michigan, NSF’s NOIRLab, the University of Nottingham, the Ohio State University, the University of Pennsylvania, the University of Portsmouth, SLAC National Accelerator Laboratory, Stanford University, the University of Sussex, and Texas A\&M University.

BASS is a key project of the Telescope Access Program (TAP), which has been funded by the National Astronomical Observatories of China, the Chinese Academy of Sciences (the Strategic Priority Research Program “The Emergence of Cosmological Structures” Grant \# XDB09000000), and the Special Fund for Astronomy from the Ministry of Finance. The BASS is also supported by the External Cooperation Program of the Chinese Academy of Sciences (Grant \# 114A11KYSB20160057), and the Chinese National Natural Science Foundation (Grant \# 12120101003, \# 11433005).

The Legacy Survey team makes use of data products from the Near-Earth Object Wide-field Infrared Survey Explorer (NEOWISE), which is a project of the Jet Propulsion Laboratory/California Institute of Technology. NEOWISE is funded by the National Aeronautics and Space Administration.

The Legacy Surveys imaging of the DESI footprint is supported by the Director, Office of Science, Office of High Energy Physics of the U.S. Department of Energy under Contract No. DE-AC02-05CH1123, by the National Energy Research Scientific Computing Center, a DOE Office of Science User Facility under the same contract; and by the U.S. National Science Foundation, Division of Astronomical Sciences under Contract No. AST-0950945 to NOAO.

\small This study made extensive use of \texttt{Python} \citep{Python3} libraries --- \textsc{\texttt{Astropy}} \citep{Astropy2013, Astropy2018, Astropy2022}, \textsc{\texttt{Numpy}} \citep{Numpy}, \textsc{\texttt{Scipy}} \citep{Scipy}, \textsc{\texttt{Matplotlib}} \citep{Matplotlib}, \textsc{\texttt{Smplotlib}} \citep{Smplotlib}, \textsc{\texttt{Pandas}} \citep{McKinney2010, Pandas}, and \textsc{\texttt{Jupyter}} \citep{Jupyter}.

\bibliographystyle{mnras}
\bibliography{Bibliography}

\appendix
\onecolumn
\counterwithin{figure}{section}
\section{Newly Identified Asymmetric Galaxies}\label{appendix:Asymmetric Galaxies}

\begin{figure*}[!h]
    \centering
        \begin{NiceTabular}{c}
          \includegraphics[width=0.60\textwidth]{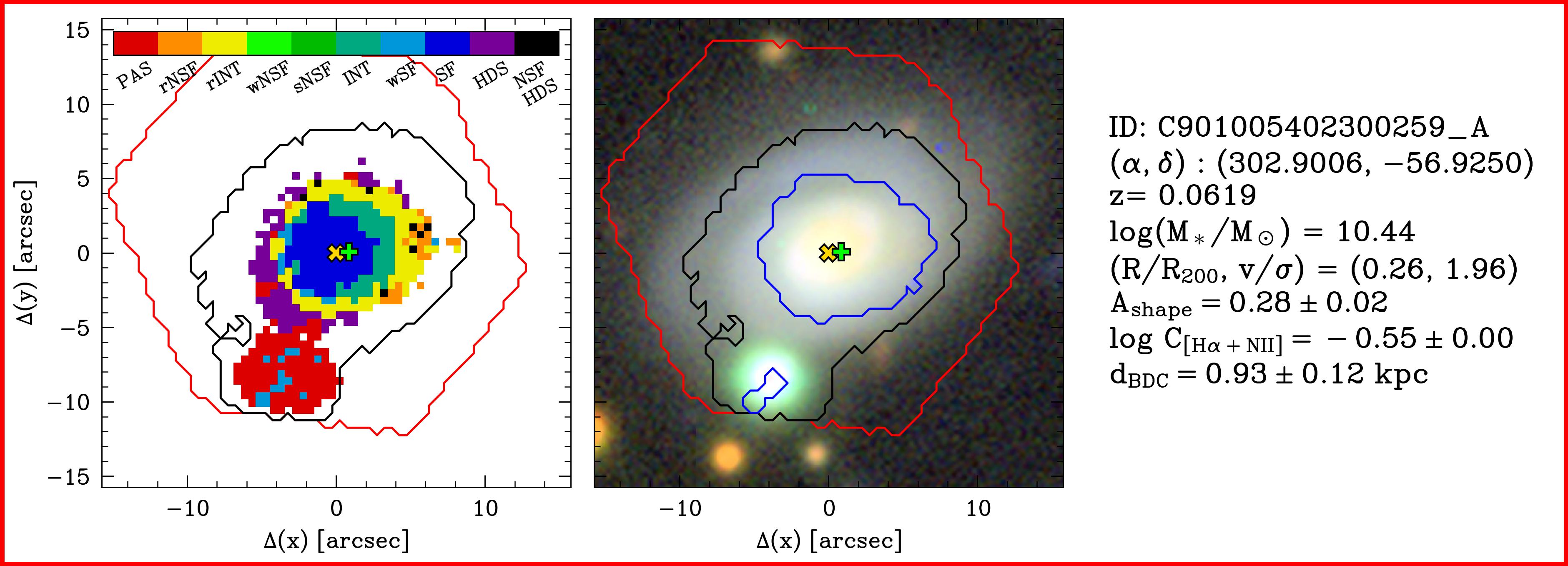}  \\
          \includegraphics[width=0.60\textwidth]{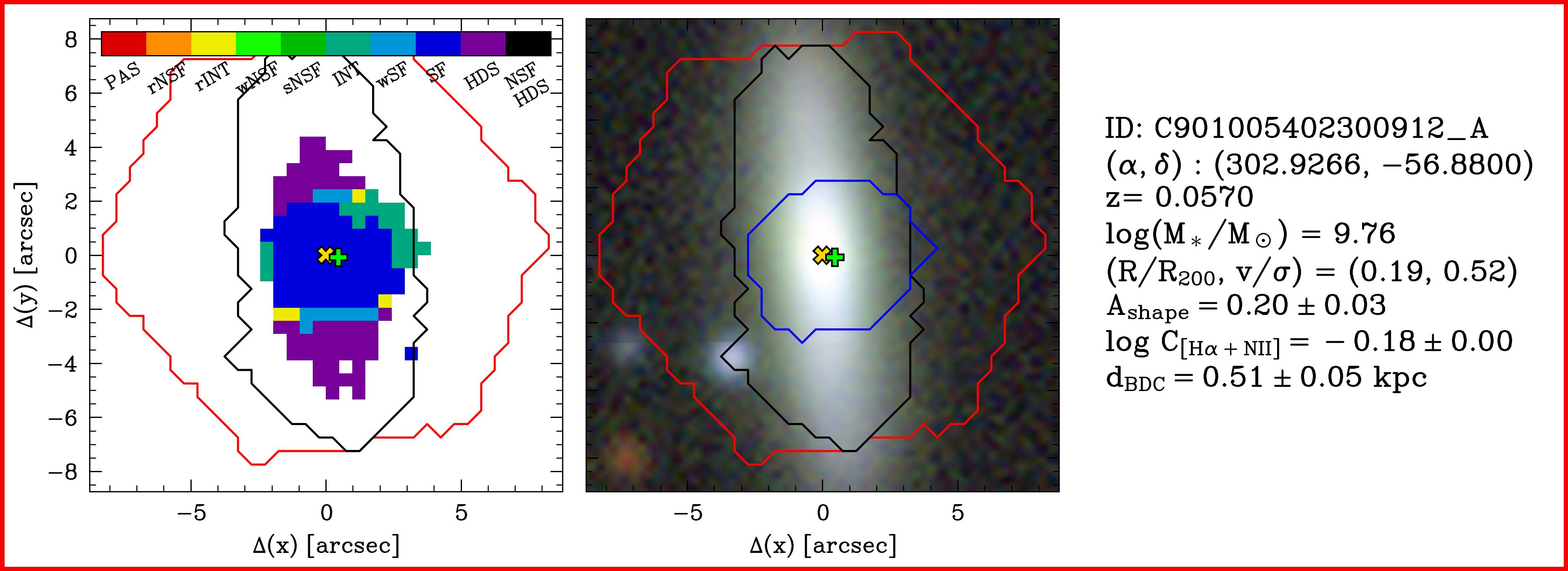}  \\
          \includegraphics[width=0.60\textwidth]{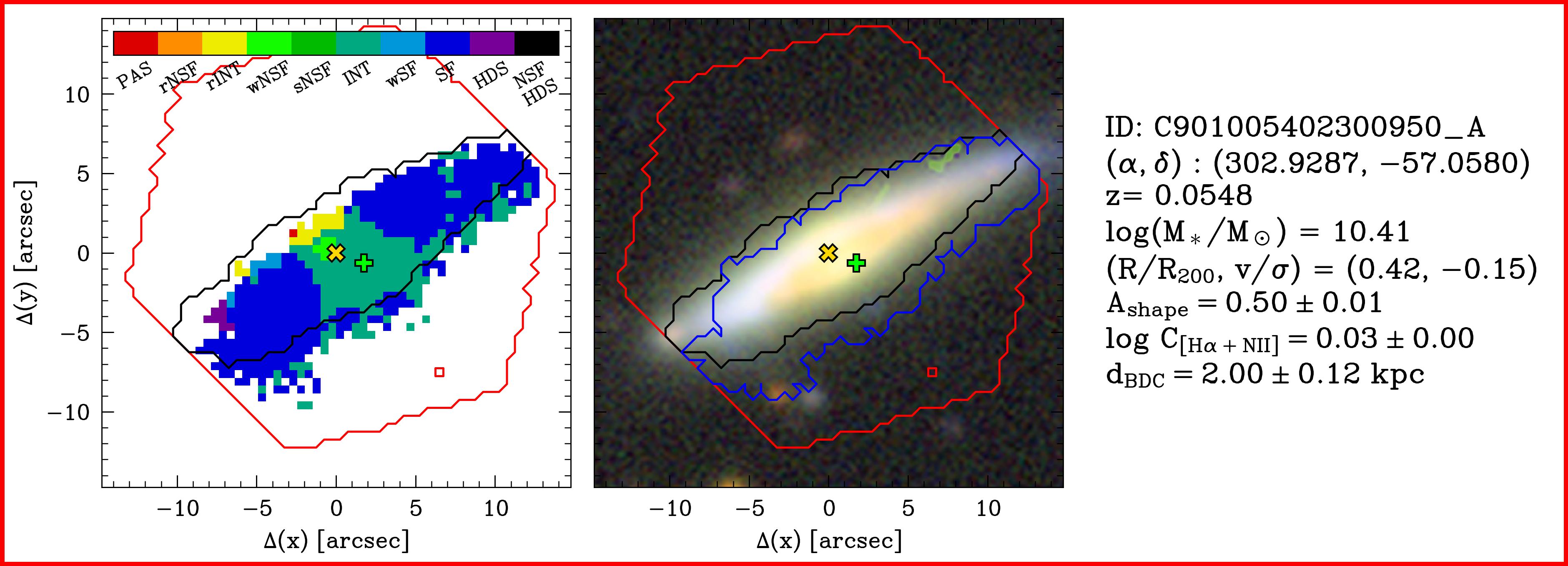}  \\        \includegraphics[width=0.60\textwidth]{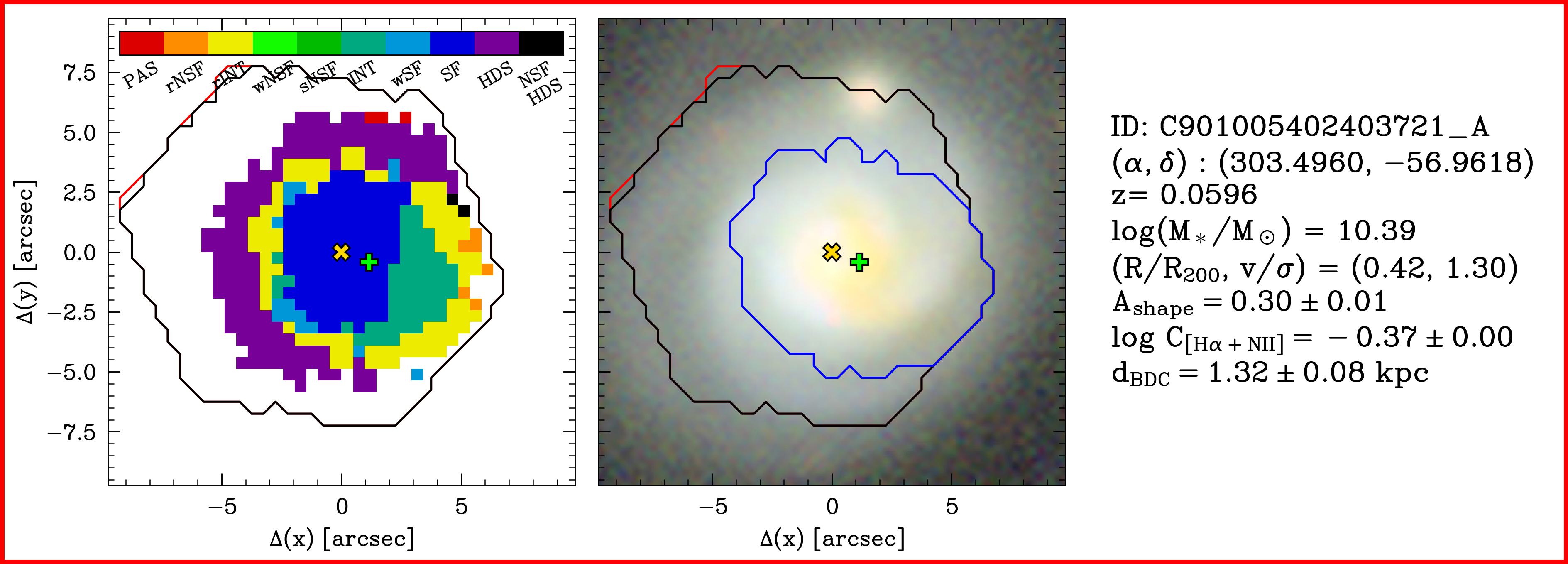} \\
          \includegraphics[width=0.60\textwidth]{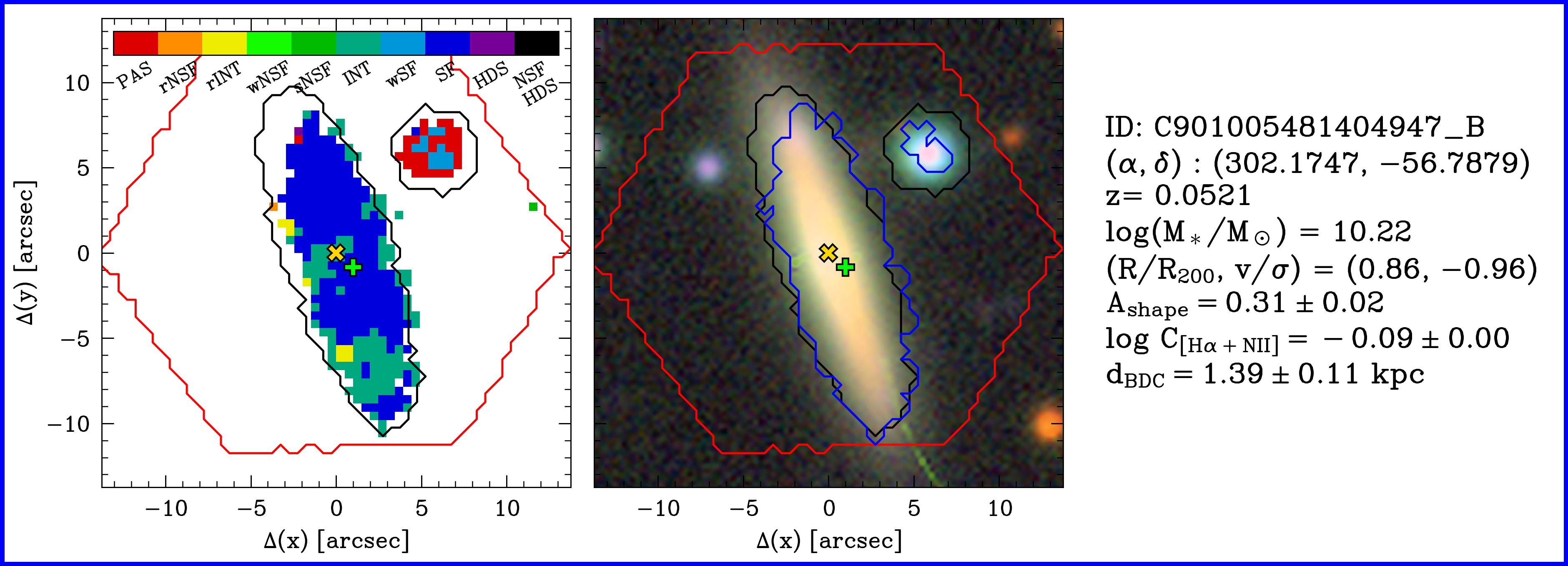}  \\
    \end{NiceTabular}
    \caption{Full compilation of newly identified asymmetric galaxies. For each galaxy, we display three panels. The left panel shows the spectroscopic classification maps as described in Section~\ref{subsection:Spectroscopic Classification}. In the middle panel, the composite $griz$ image from the Legacy Survey \citep{Dey2019} is displayed. The blue contour traces the ionised gas distribution. For both the left and middle panels, the red and black contours mark the hexabundle edges and the stellar continuum around $\rm H\alpha$ and $\rm [NII]\lambda6584$ at $\rm SNR =2$, respectively. The yellow cross and the cyan plus show the galaxy and the ionised gas binary detection centre, respectively. The right panel lists the properties of the galaxy---\textit{ID, RA, DEC, redshift, stellar mass, cluster-centric distance, peculiar velocity, shape asymmetry, concentration, and offset between galaxy and ionised gas binary detection centres}. The frame colour implies whether the galaxy is associated with the NW (blue) or the MC (red) asymmetric galaxies; otherwise, it is black.}
    \label{fig:Asymmetric Galaxies}
\end{figure*}

\begin{figure*}[!h]
    \centering
        \begin{NiceTabular}{c}
          \includegraphics[width=0.60\textwidth]{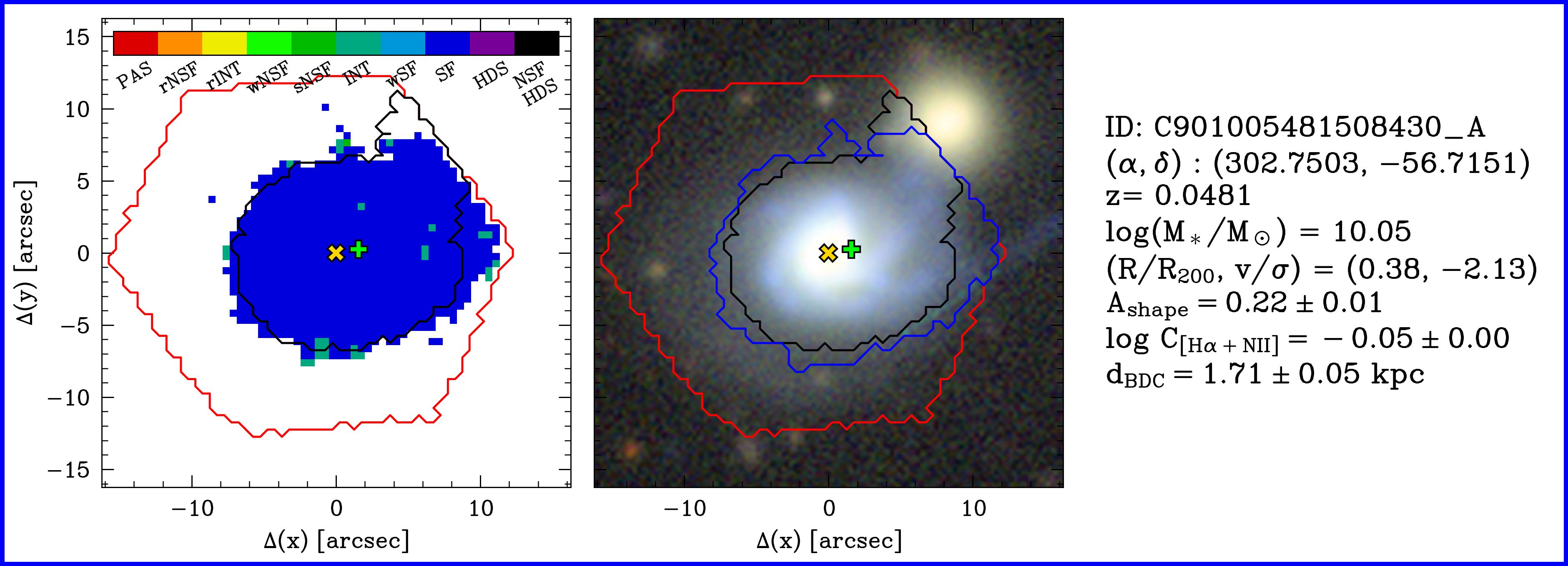}   \\ 
          \includegraphics[width=0.60\textwidth]{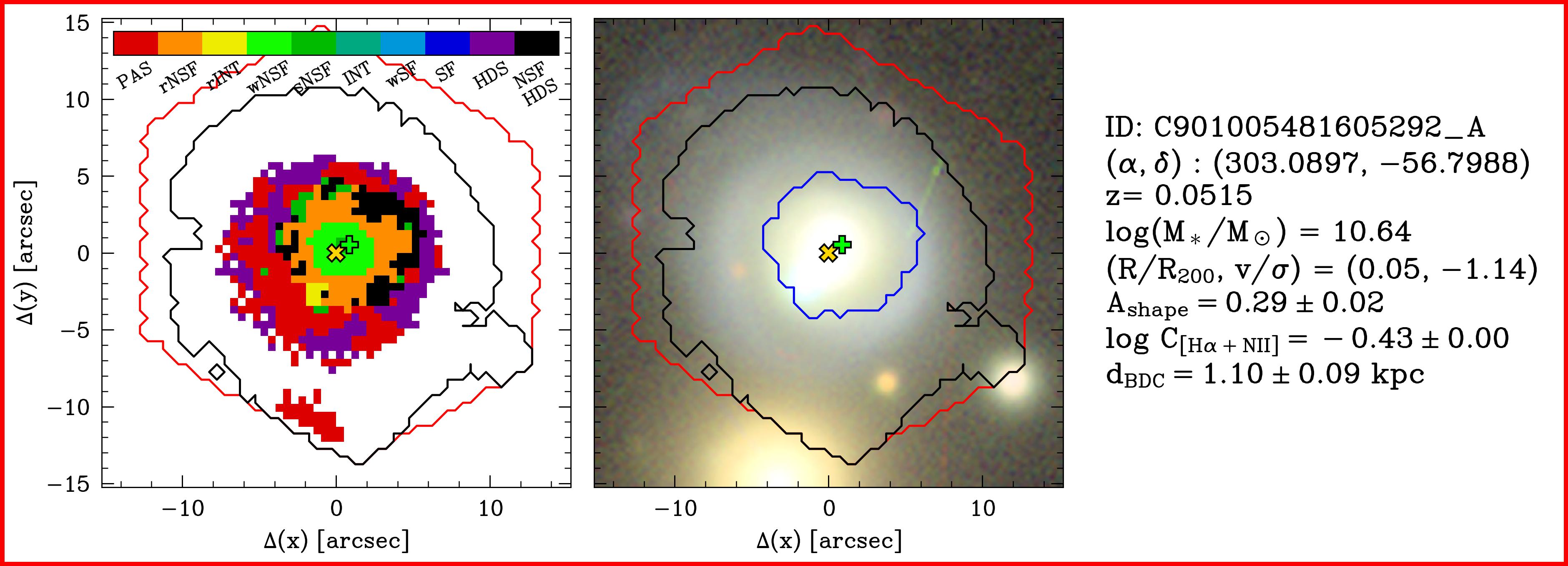}   \\
          \includegraphics[width=0.60\textwidth]{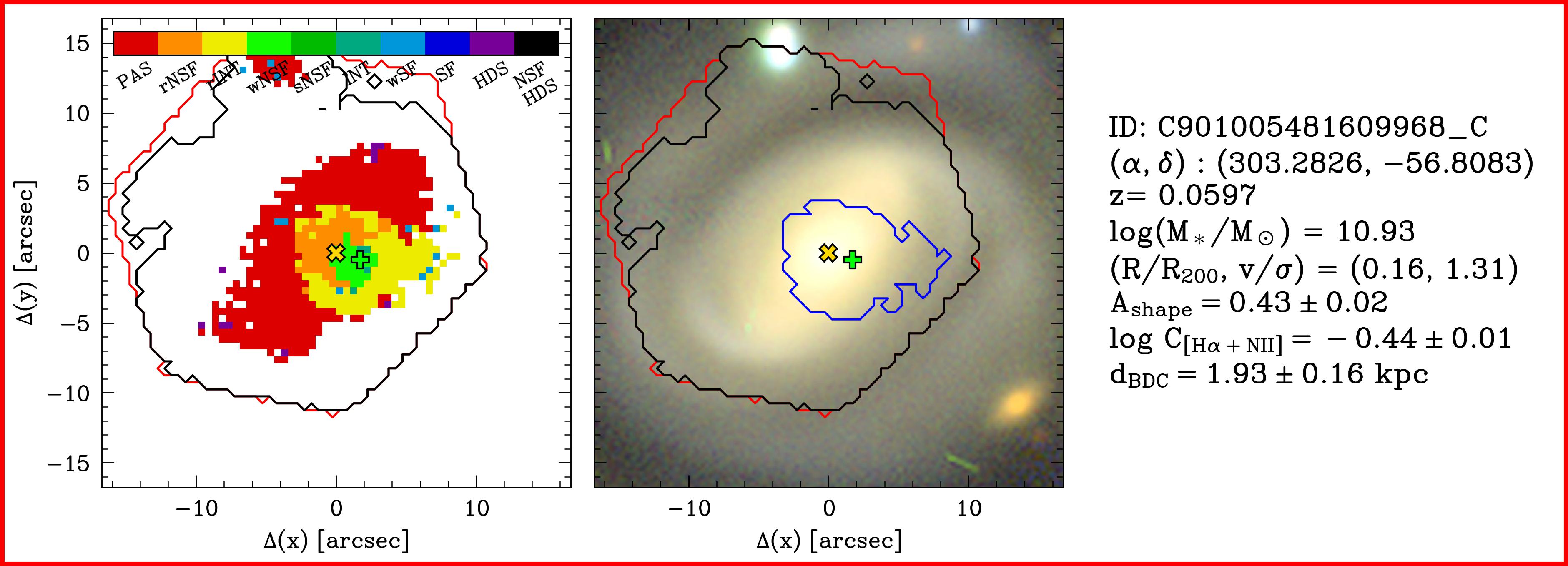}   \\
          \includegraphics[width=0.60\textwidth]{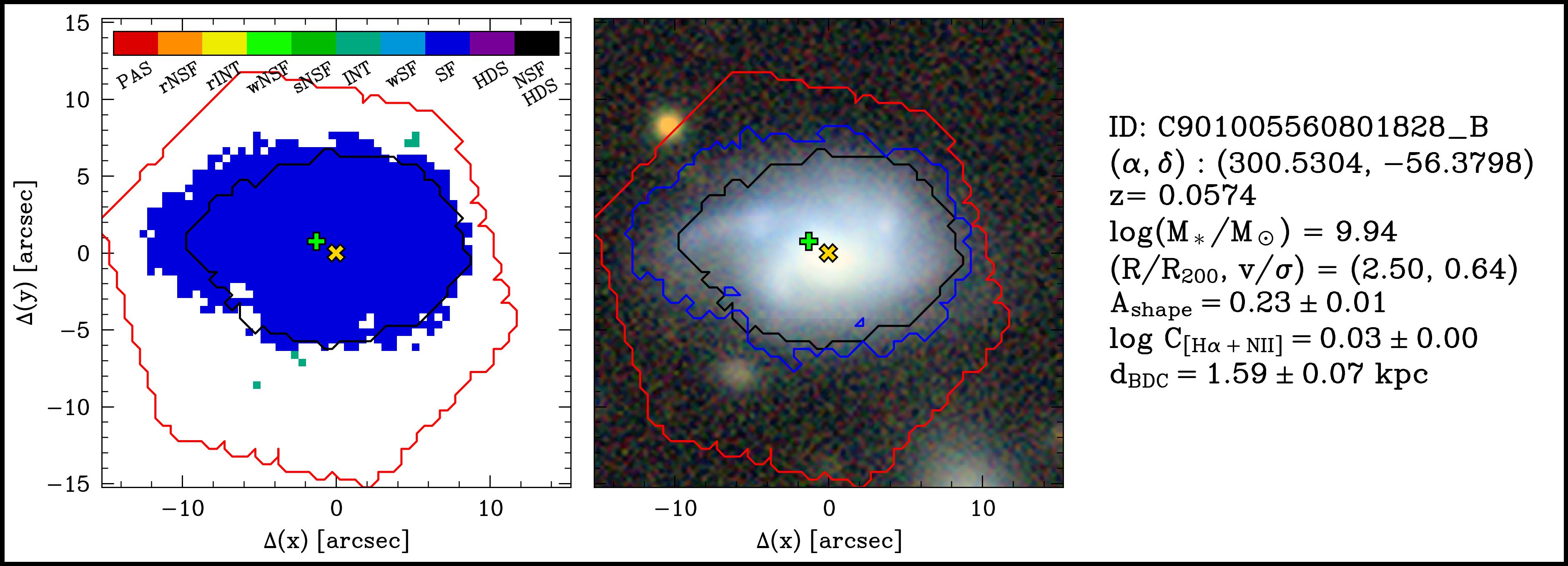}   \\
          \includegraphics[width=0.60\textwidth]{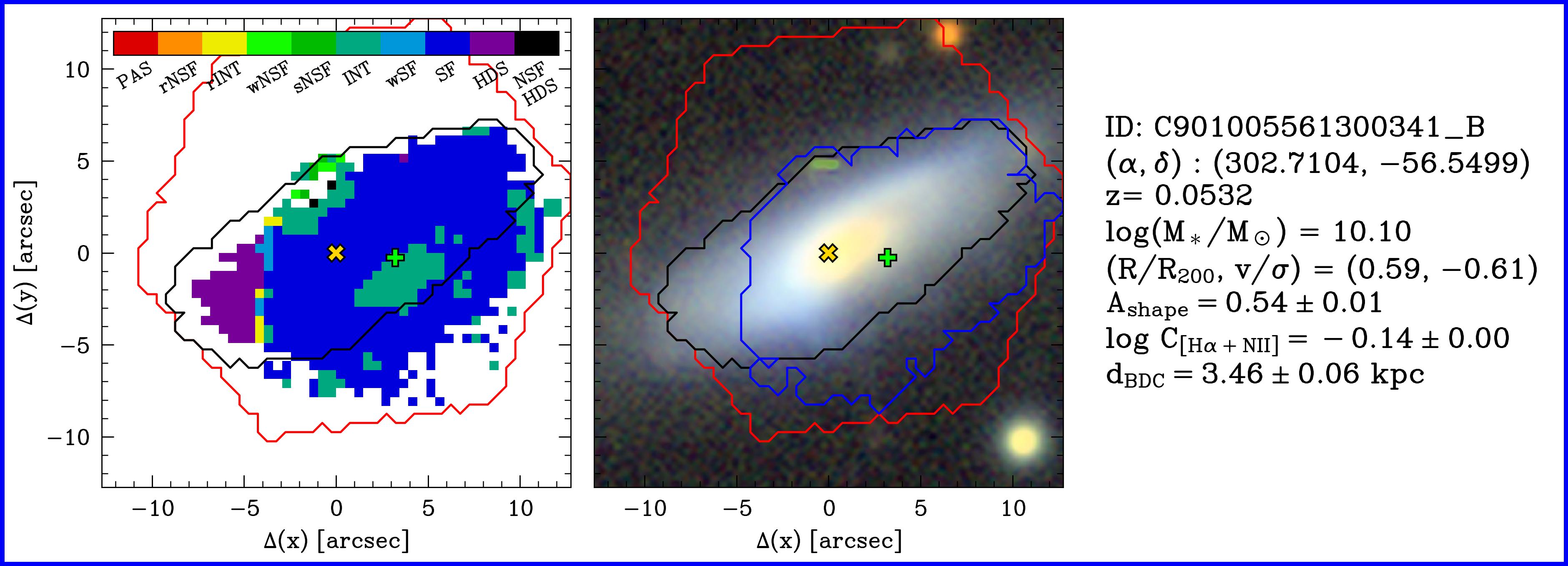}
    \end{NiceTabular}
    \caption{(Continued).}
    \label{fig:Asymmetric Galaxies}
\end{figure*}

\clearpage

\begin{figure*}[!h]
    \centering
        \begin{NiceTabular}{c}
          \includegraphics[width=0.60\textwidth]{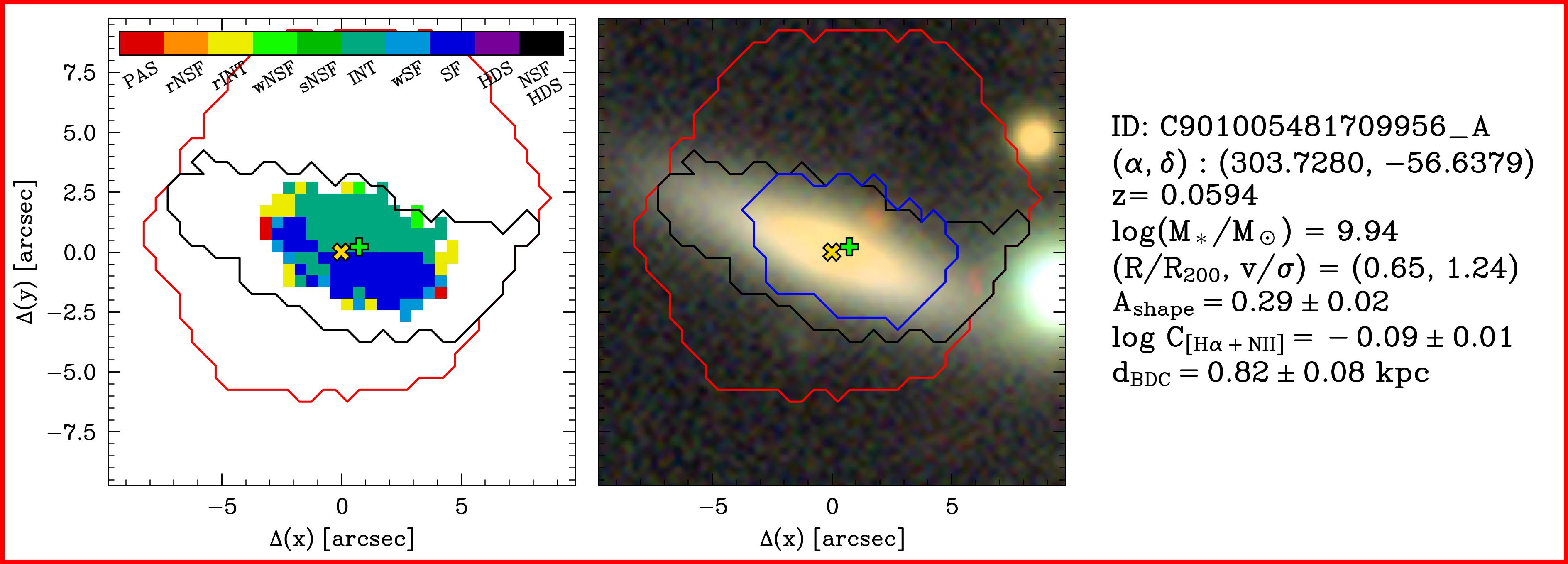}   \\ 
          \includegraphics[width=0.60\textwidth]{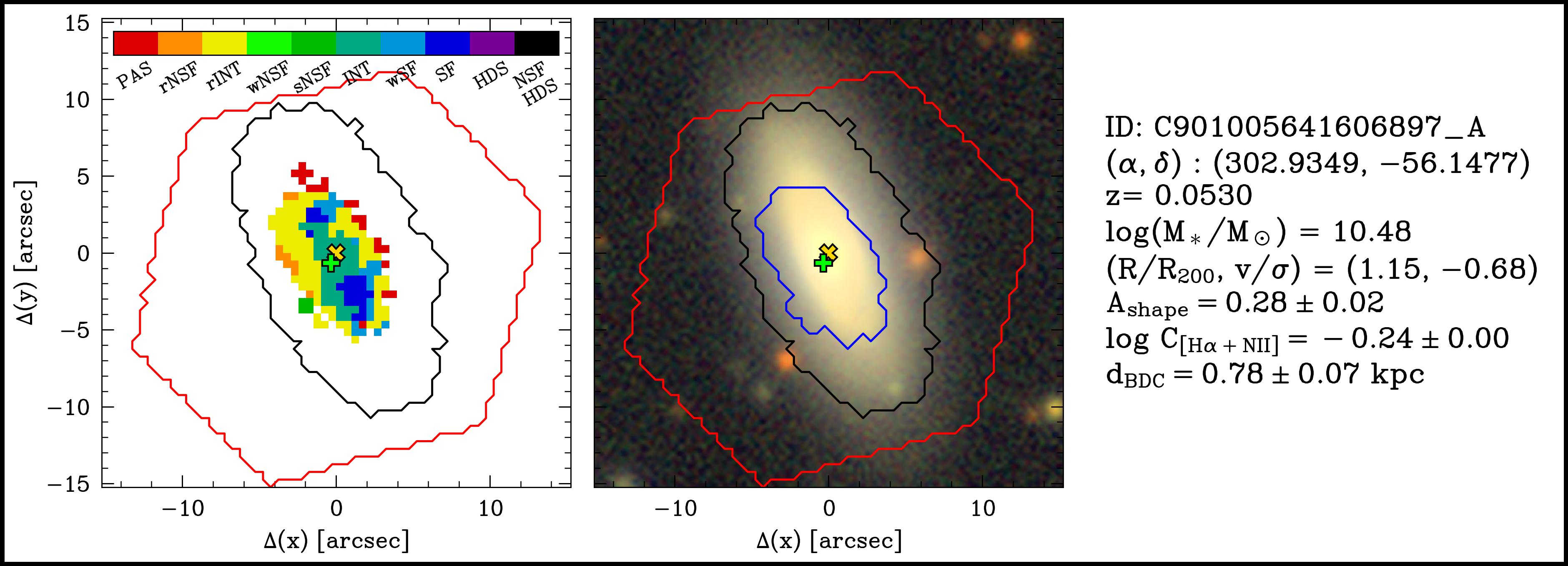} 
        \end{NiceTabular}
    \caption{(Continued).}
    \label{fig:Asymmetric Galaxies}
\end{figure*}

\section{Jellyfish Galaxies}\label{appendix:Jellyfish Galaxies}
\begin{figure*}[!h]
    \centering
        \begin{tabular}{ccc}
          \includegraphics[width=0.3\textwidth]{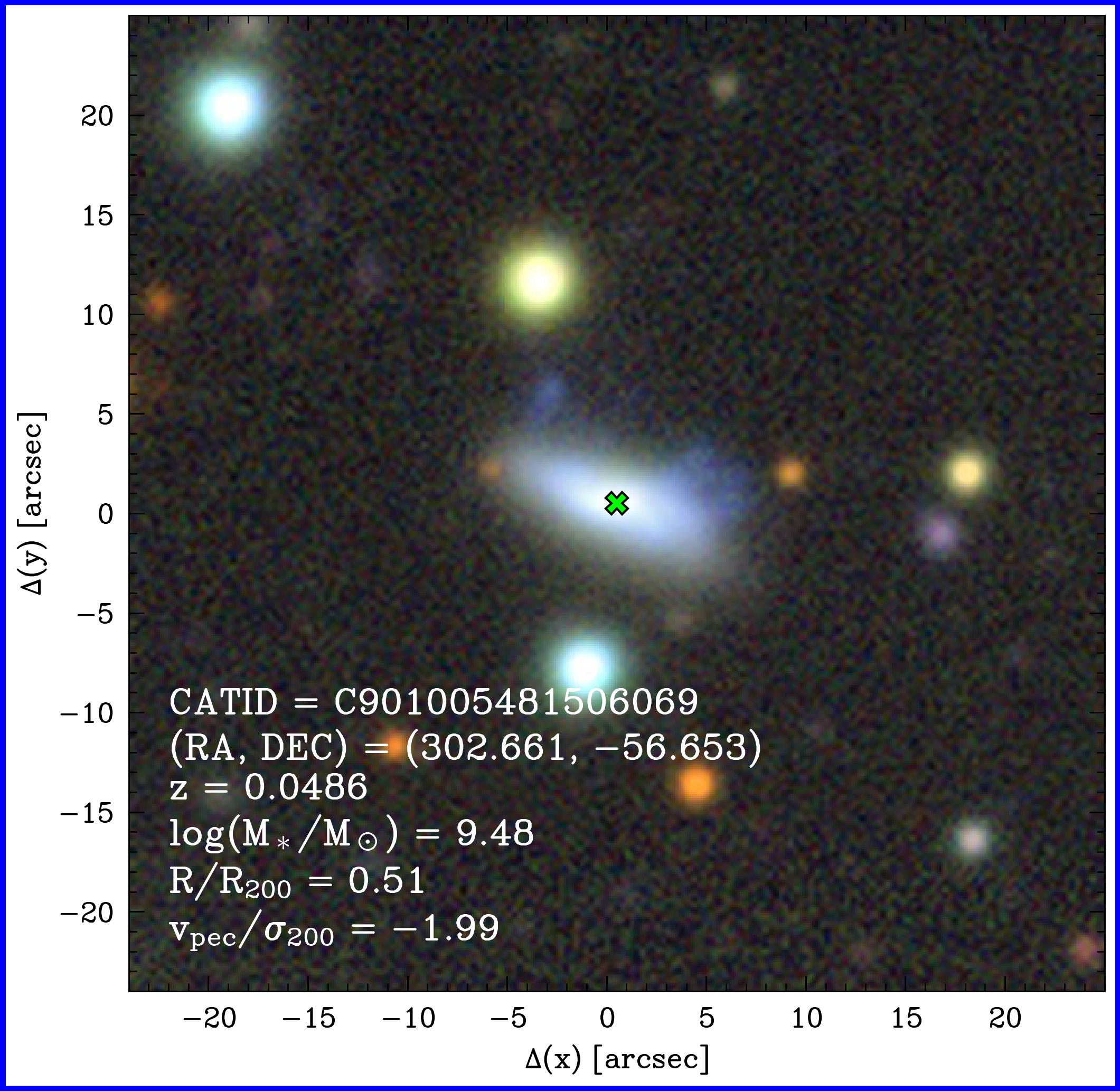} & \includegraphics[width=0.3\textwidth]{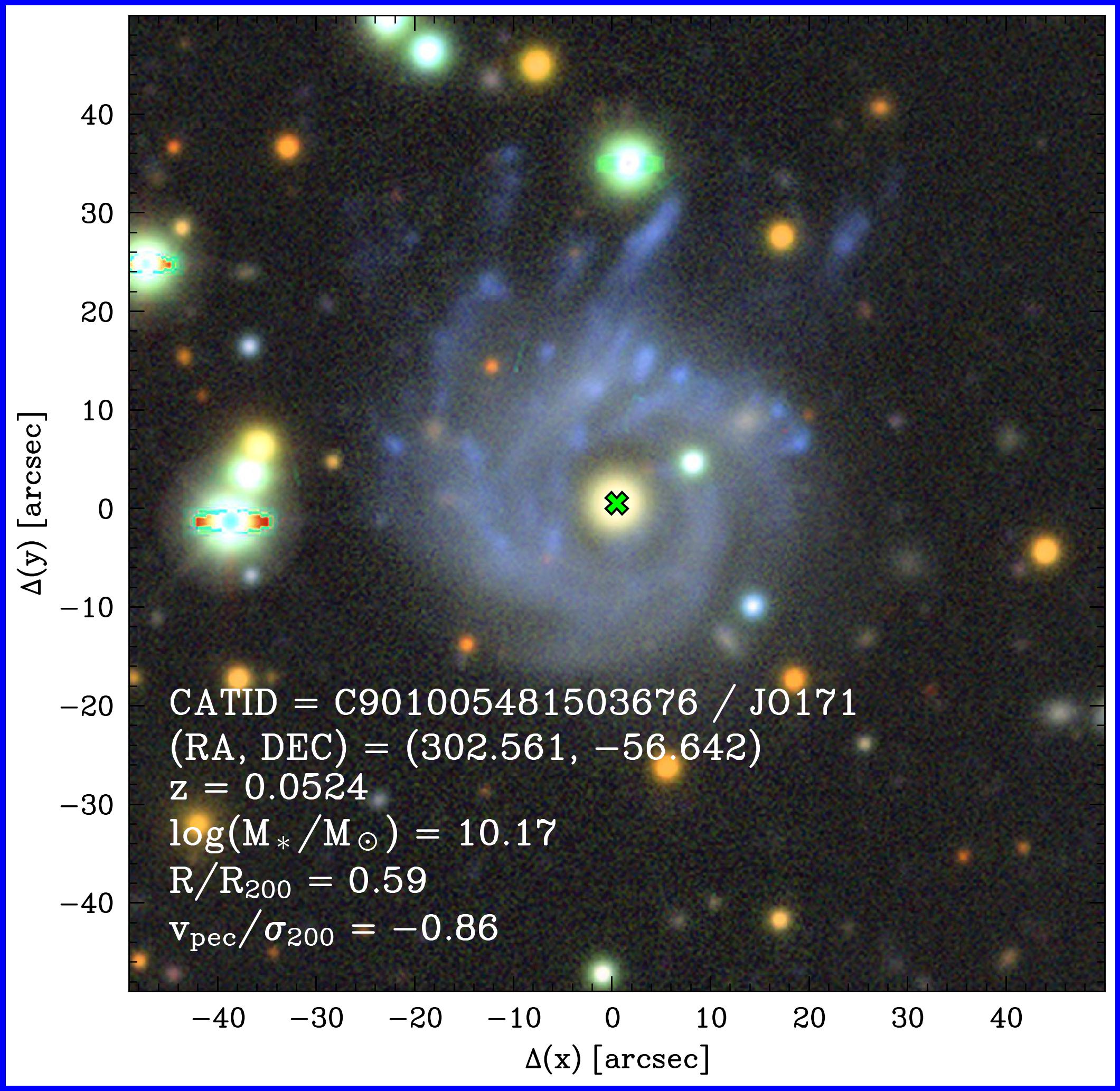} &           \includegraphics[width=0.305\textwidth]{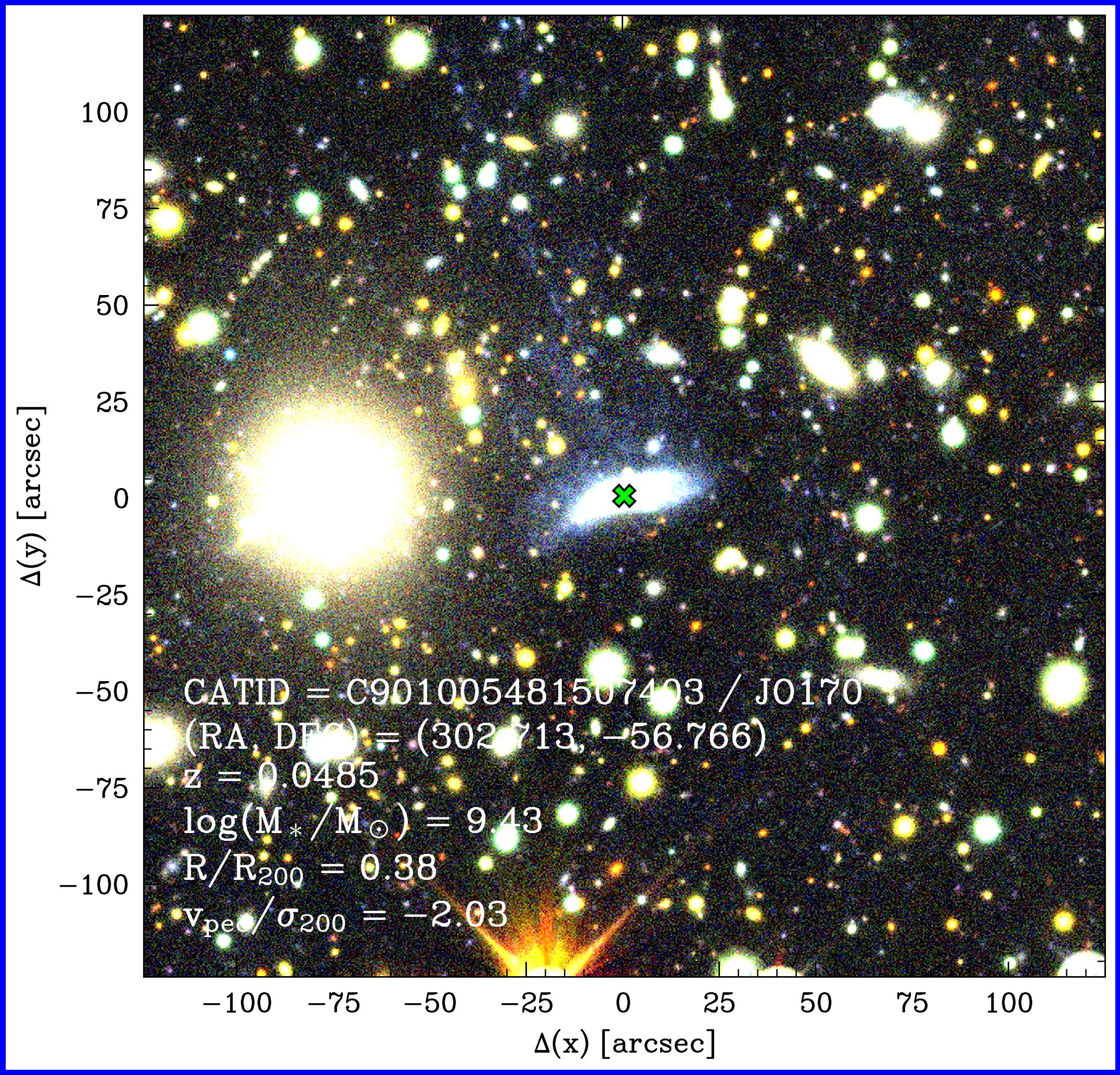}
        \end{tabular}
    \caption{The composite $griz$ images from Legacy Survey \citep{Dey2019} of the jellyfish galaxy sample. The left one is identified by our visual inspection, while others have been identified by the GASP survey \citep[\texttt{JO170} and \texttt{JO171};][]{Poggianti2016}. The blue frames means the same as in Figure~\ref{fig:Asymmetric Galaxies}.}
    \label{fig:Jellyfish Galaxies}
\end{figure*}

\clearpage

\section{New Identified Truncated Galaxies}\label{appendix:Truncated Galaxies}
\begin{figure*}[!h]
    \centering
        \begin{tabular}{c}
          \includegraphics[width=0.6\textwidth]{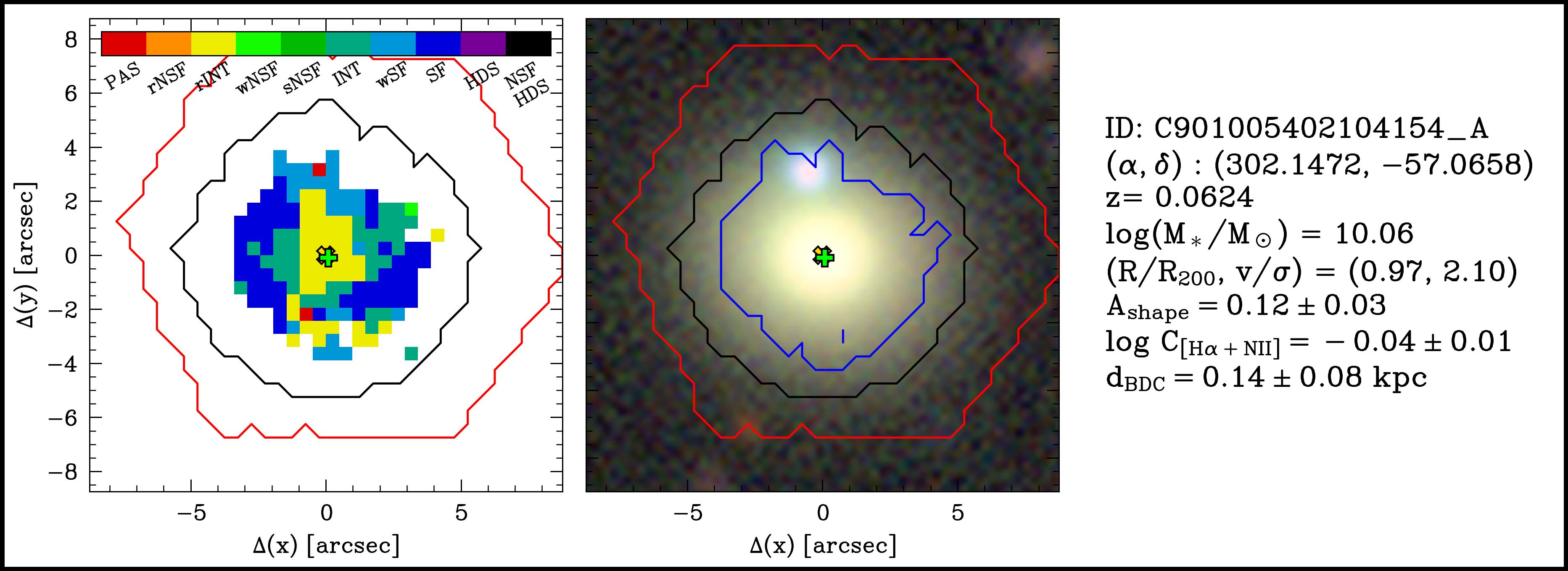}   \\ 
          \includegraphics[width=0.6\textwidth]{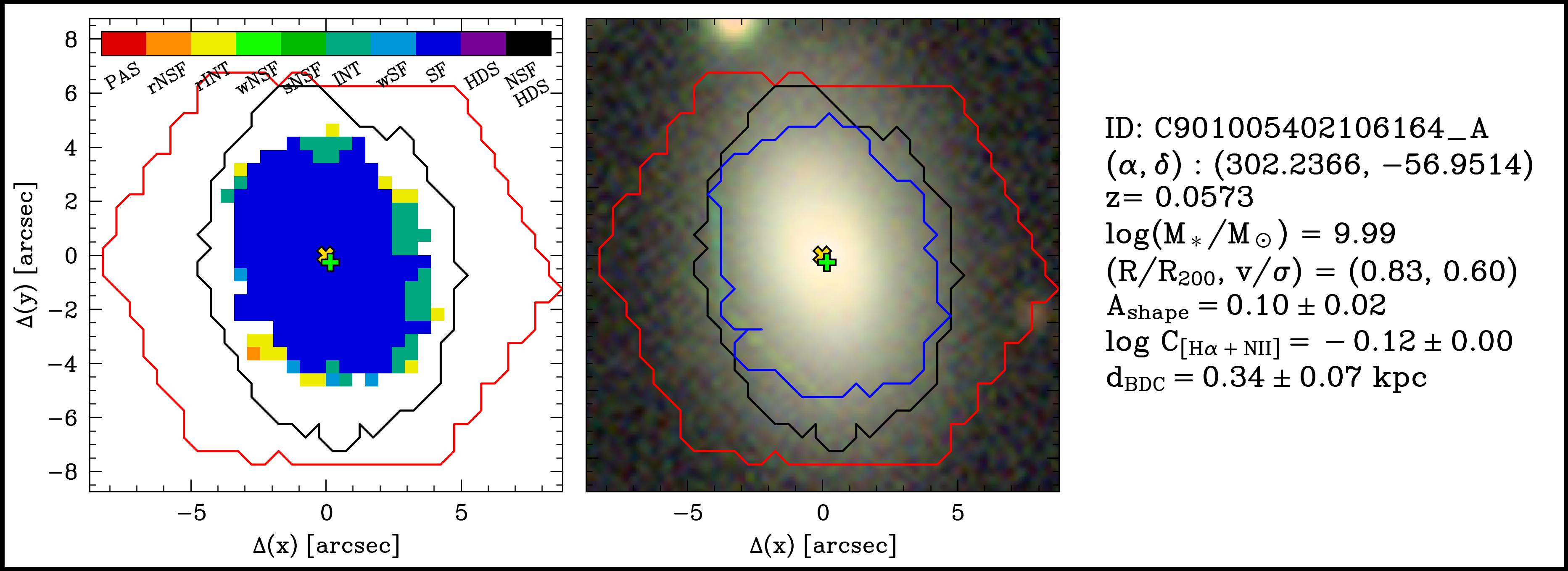}   \\
          \includegraphics[width=0.6\textwidth]{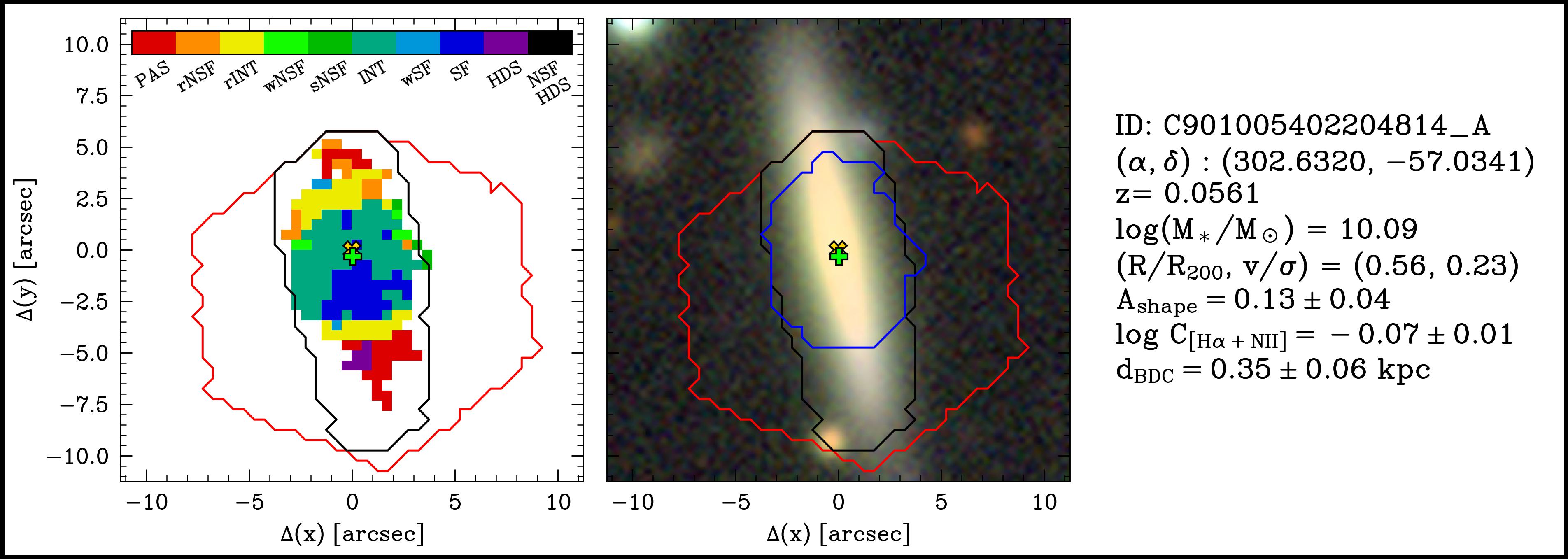}   \\
          \includegraphics[width=0.6\textwidth]{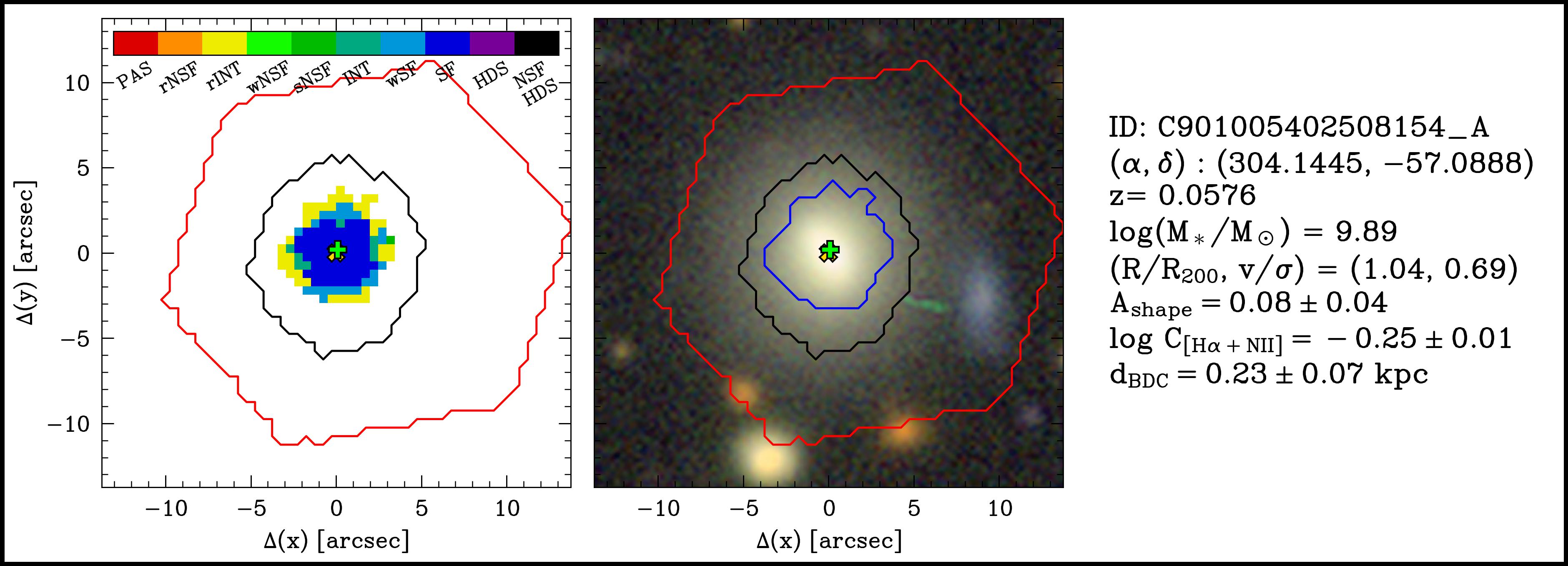}   \\
          \includegraphics[width=0.6\textwidth]{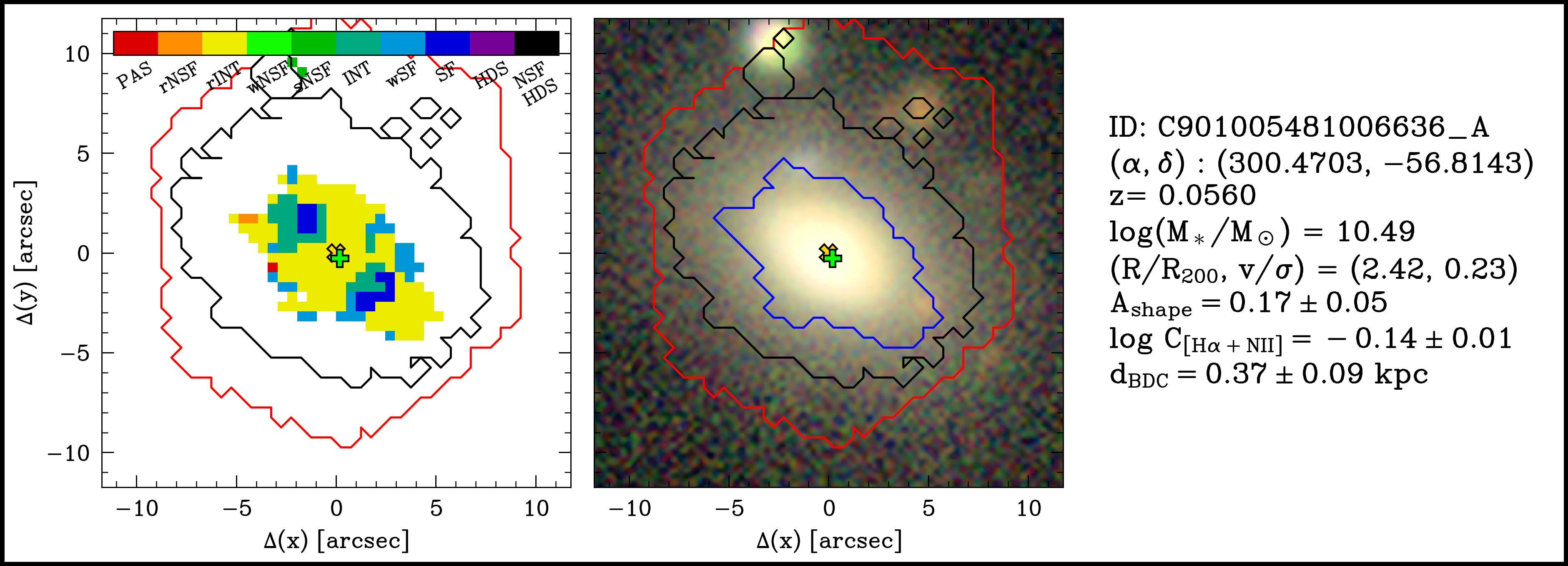}          
    \end{tabular}
    \caption{Same as Figure~\ref{fig:Asymmetric Galaxies}, but for truncated galaxies.}
    \label{fig:Truncated Galaxies}
\end{figure*}

\begin{figure*}[!h]
    \centering
        \begin{tabular}{c}
            \includegraphics[width=0.6\textwidth]{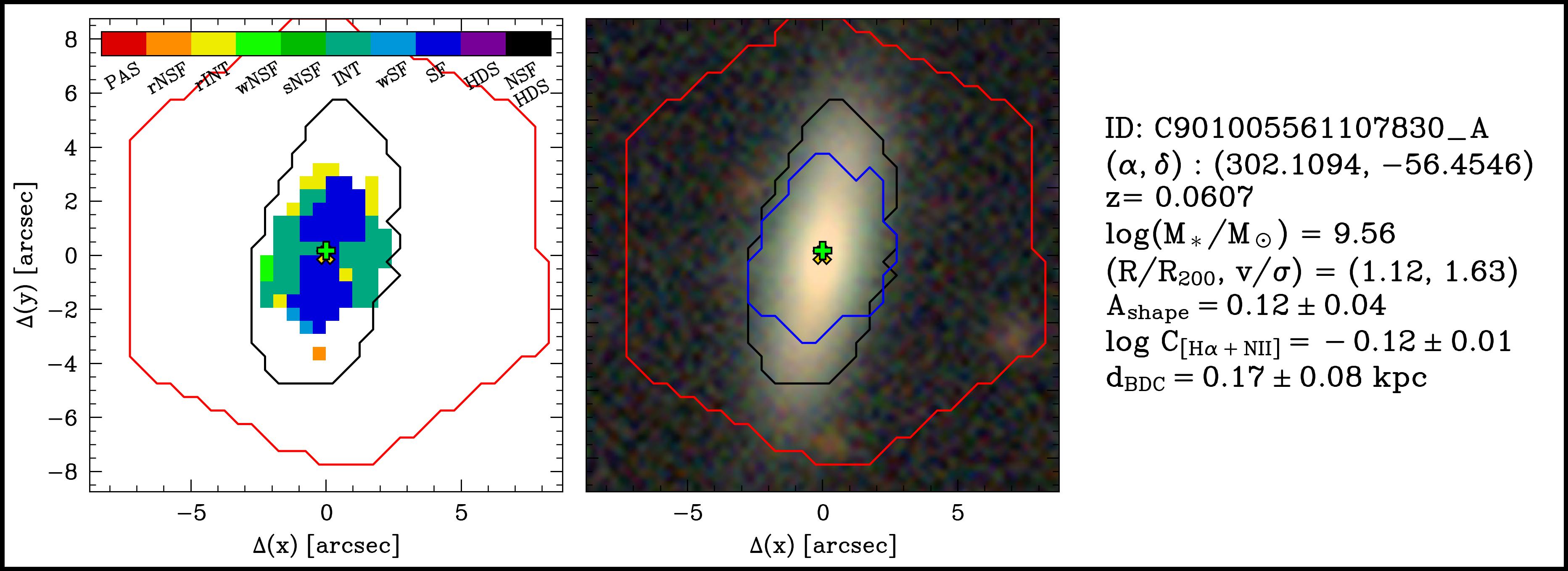}   \\
              \includegraphics[width=0.6\textwidth]{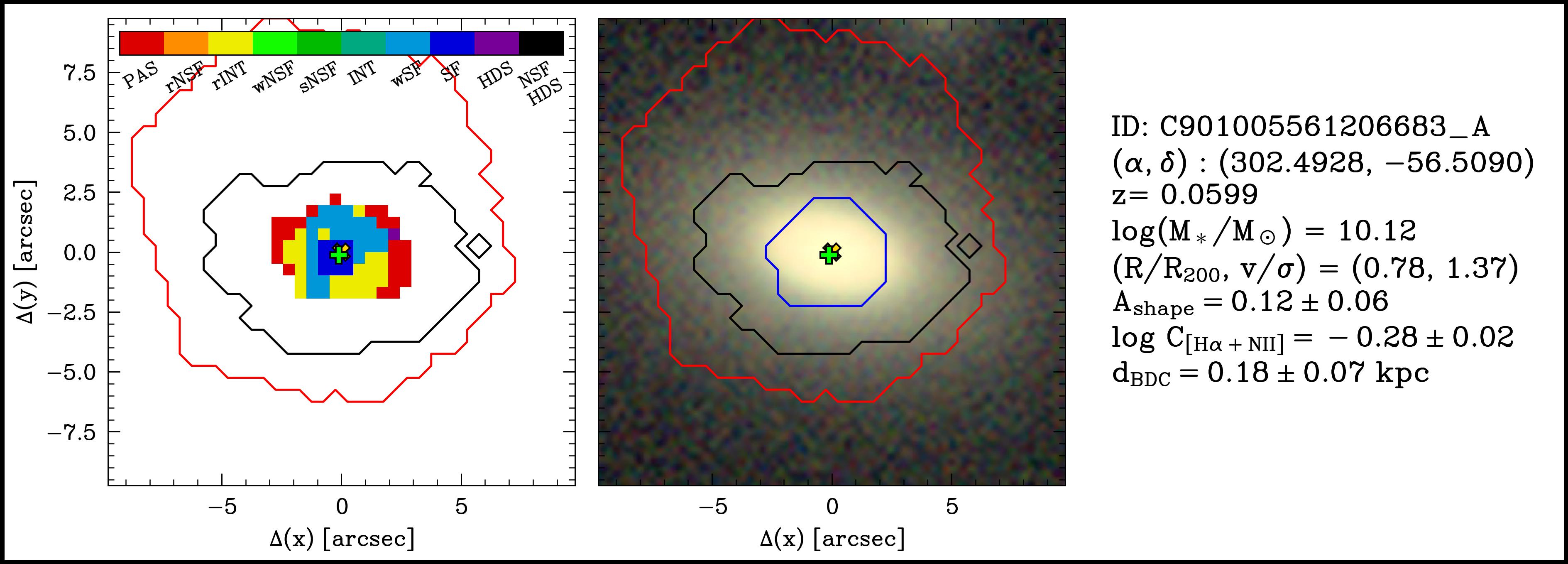}  \\
              \includegraphics[width=0.6\textwidth]{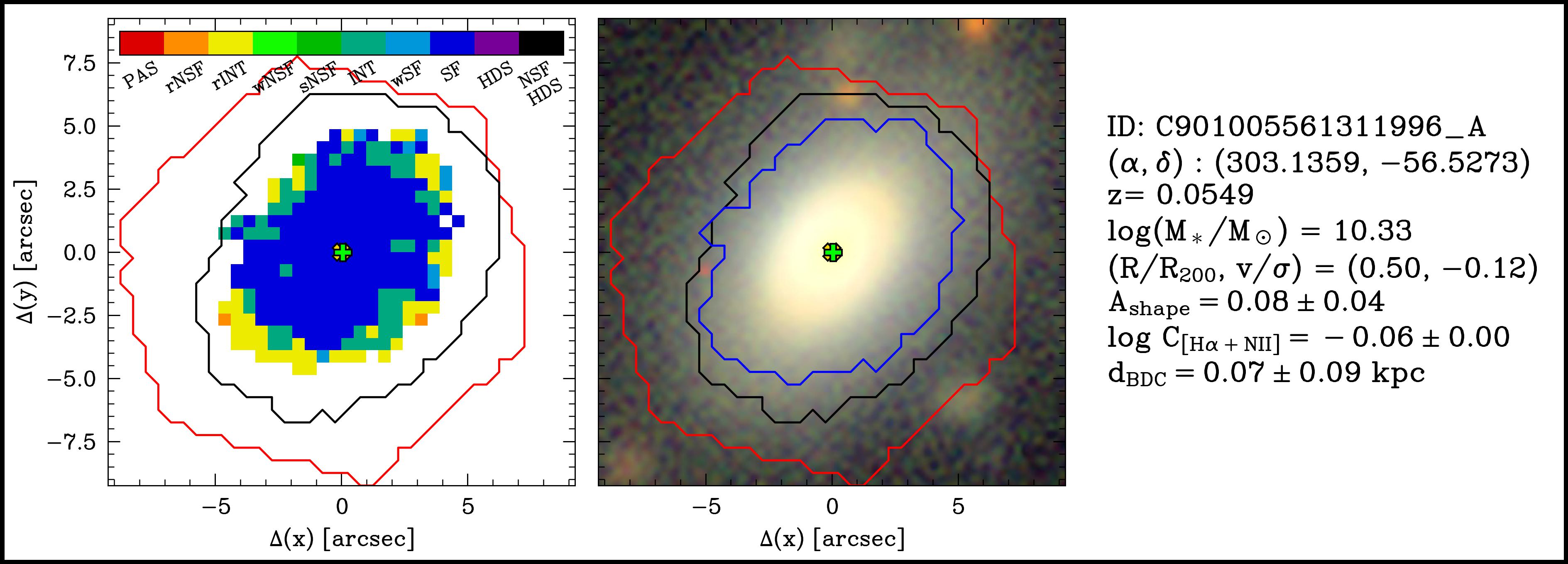} \\
              \includegraphics[width=0.6\textwidth]{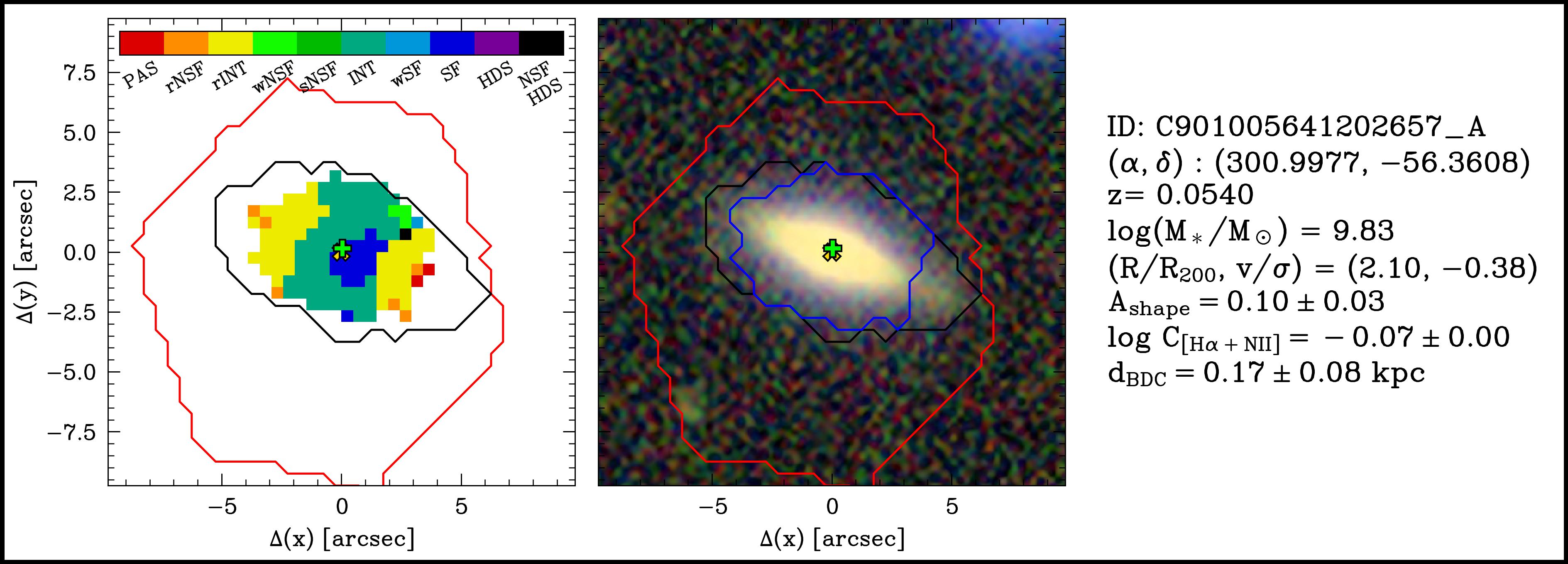} \\
              \includegraphics[width=0.6\textwidth]{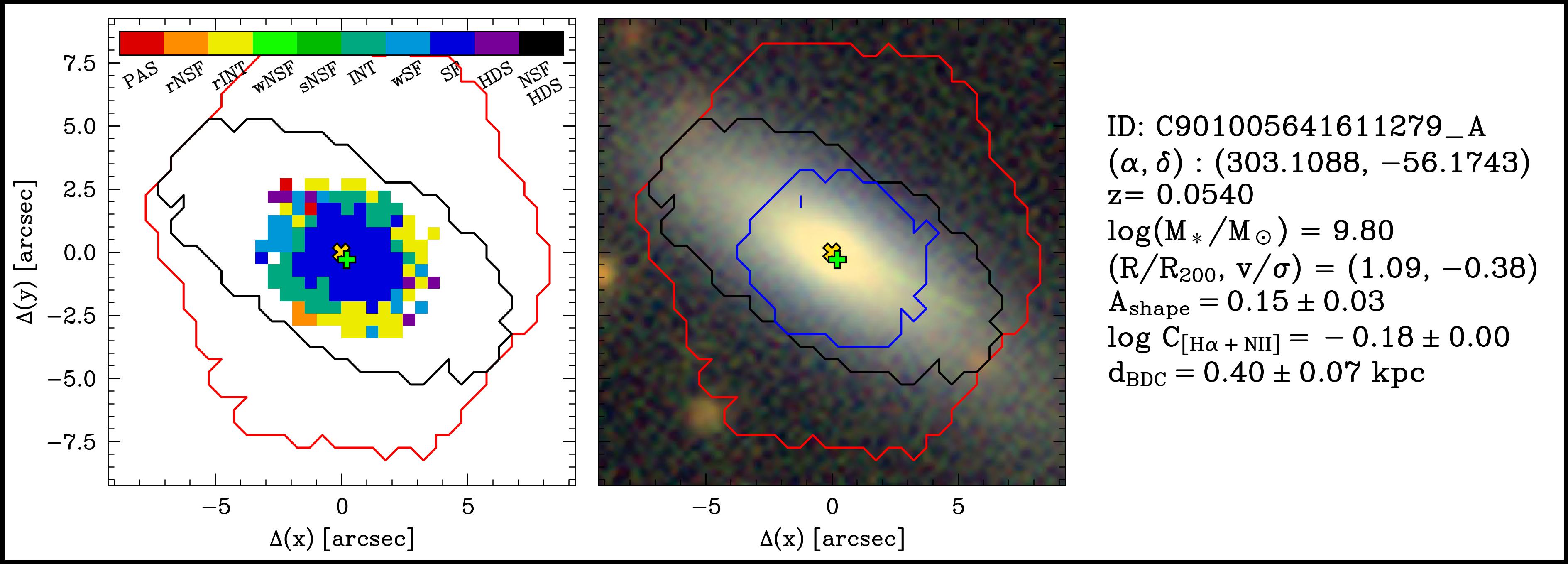} 
    \end{tabular}
    \caption{(Continued).}
\end{figure*}

\clearpage

\begin{figure*}[!h]
    \centering
        \begin{tabular}{c}
            \includegraphics[width=0.6\textwidth]{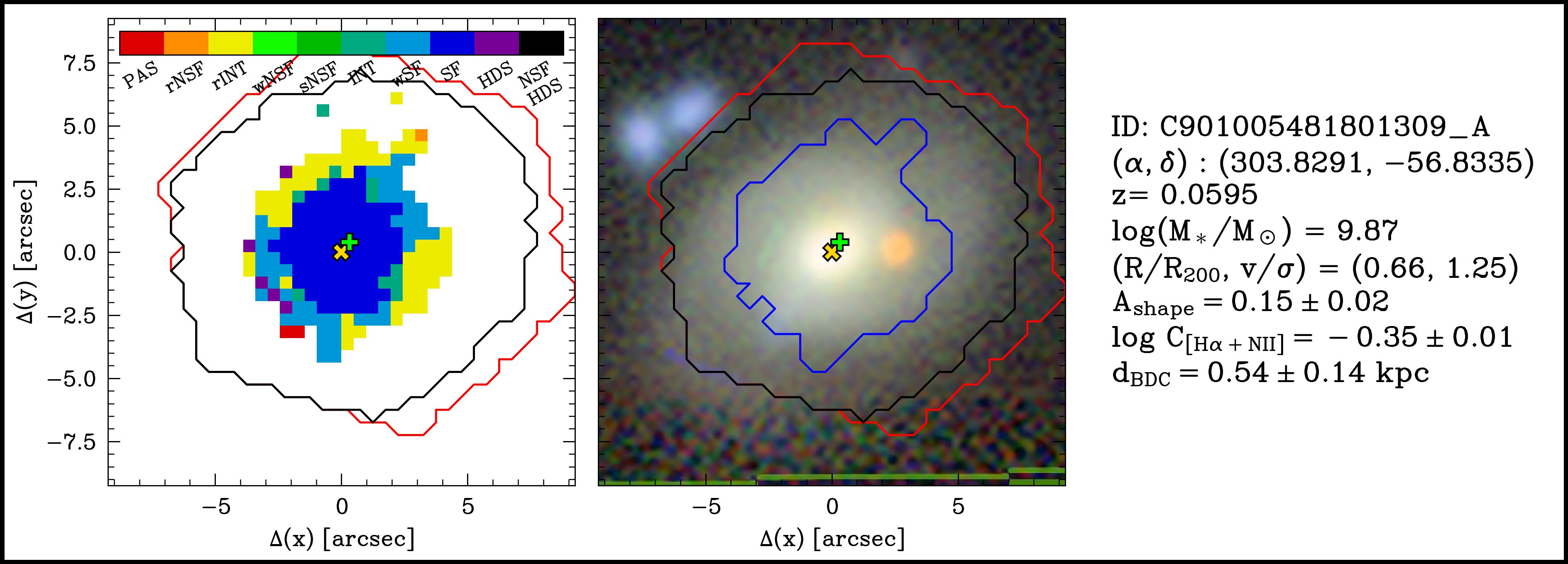} \\
            \includegraphics[width=0.6\textwidth]{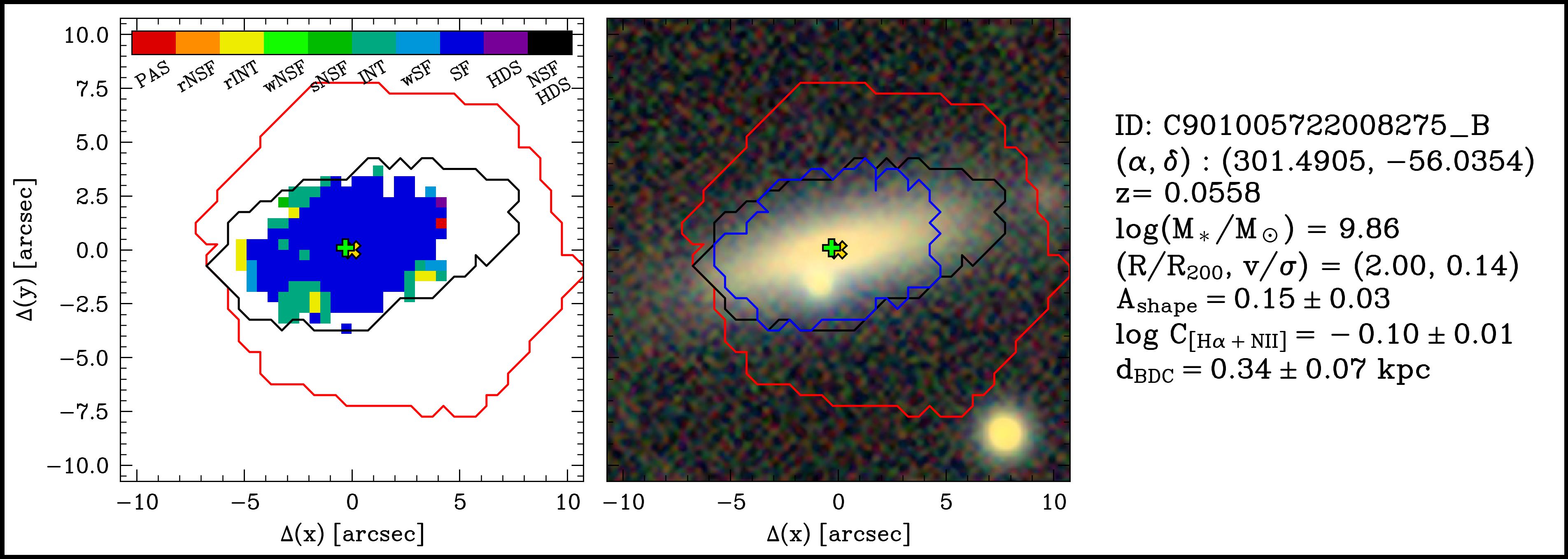}
    \end{tabular}
    \caption{(Continued).}
\end{figure*}

\section{Fully quenched $\rm H\delta-$strong galaxy}\label{appendix:HDS-q}
\begin{figure*}[!h]
    \centering
    \includegraphics[width=0.75\textwidth]{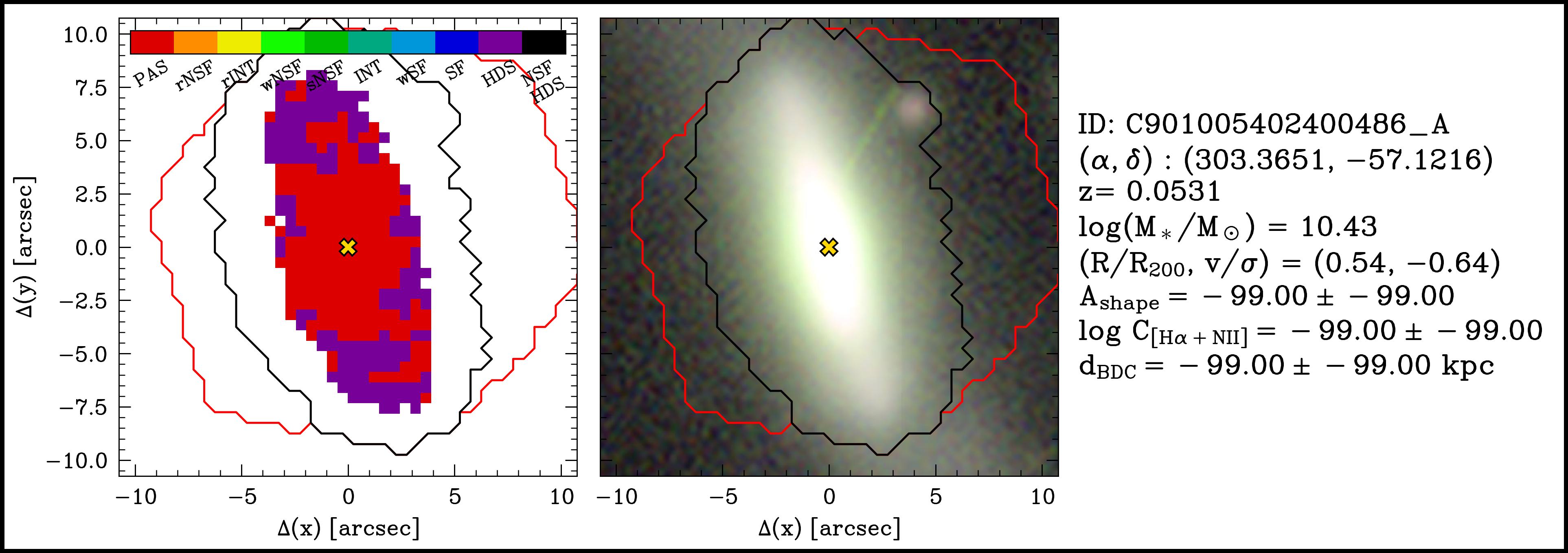}
    \caption{Same as Figure~\ref{fig:Asymmetric Galaxies}, but for the fully quenched H$\delta-$strong galaxy \texttt{C901005402400486}.}
\end{figure*}

\section{Formulation used to determine tail orientations}\label{appendix:Formulas}
We define three different vectors - tail vector ($\vec{\rm T}$), cluster-centric vector ($\vec{\rm R}$), and merger axis vector ($\vec{\rm M}$), as follows

\begin{equation}
\vec{\mathrm{T}} = (x_{\mathrm{gas}} - x_{\mathrm{galaxy}},\ y_{\mathrm{gas}} - y_{\mathrm{galaxy}})
\end{equation}

\[\vec{\mathrm{R}} = (x_{\mathrm{cluster}} - x_{\mathrm{galaxy}},\ y_{\mathrm{cluster}} - y_{\mathrm{galaxy}})\]

\[\vec{\mathrm{M}} = (x_{\mathrm{BCG,\ main}} - x_{\mathrm{BCG,\ NW}},\ y_{\mathrm{BCG, \ main}} - y_{\mathrm{BCG, \ NW}})\]

\noindent Here, $(x_{\mathrm{cluster}}, \ y_{\mathrm{cluster}})$ represents the cluster centre and $(x_{\mathrm{galaxy}}, \ y_{\mathrm{galaxy}})$ is the galaxy centre. The centroid of the ionised gas distribution, ($x_{\mathrm{gas}}, y_{\mathrm{gas}}$), is defined as follows:

\begin{equation}\label{eq:Gas centroid}
(x_{\mathrm{gas}}, y_{\mathrm{gas}}) = \biggl( \frac{\sum_{i=1}^{N_{spaxel}} x_i \times w_i}{\sum_{i=1}^{N_{spaxel}} w_i} , \frac{\sum_{i=1}^{N_{spaxel}} y_i \times w_i}{\sum_{i=1}^{N_{spaxel}} w_i} \biggl)
\end{equation}

\noindent where the weight $w_i$ defined as 1 if there is emission detection for the $\text{spaxel}(x_i, y_i)$, and 0 otherwise. We then determine the angle of the tail relative to the cluster centre ($\theta_{\mathrm{tail,c}}$) and relative to the merger axis ($\theta_{\mathrm{tail,m}}$) using the formula below

\begin{equation}
\mathrm{cos}(\theta_{\mathrm{tail,c}}) = \frac{\vec{\mathrm{T}} \cdot \vec{\mathrm{R}}}{|\vec{\mathrm{T}}|\ |\vec{\mathrm{R}}|} 
\end{equation}

\begin{equation} 
\mathrm{cos}(\theta_{\mathrm{tail,m}}) = \frac{\vec{\mathrm{T}} \cdot \vec{\mathrm{M}}}{|\vec{\mathrm{T}}|\ |\vec{\mathrm{M}}|} 
\end{equation}

\noindent In order to estimate the absolute misalignments between tails and the merger axis, we also applied the following

\begin{equation}
    \theta_{\mathrm{tail,m}} = \mathrm{min}(\theta_{\mathrm{tail,m}}, \ 180-\theta_{\mathrm{tail,m}})
\end{equation}

\vskip 5mm

\noindent The directional (i.e., N, E, S, W) orientations are defined as below
\begin{equation}\label{eq:Tail direction}
\mathrm{tan}(\theta_{\mathrm{tail,W}}) = \frac{y_{\mathrm{gas}} - y_{galaxy}}{x_{\mathrm{gas}} - x_{\mathrm{galaxy}}}
\end{equation}

\begin{figure*}[!hb]
    \centering
    \includegraphics[width=\textwidth]{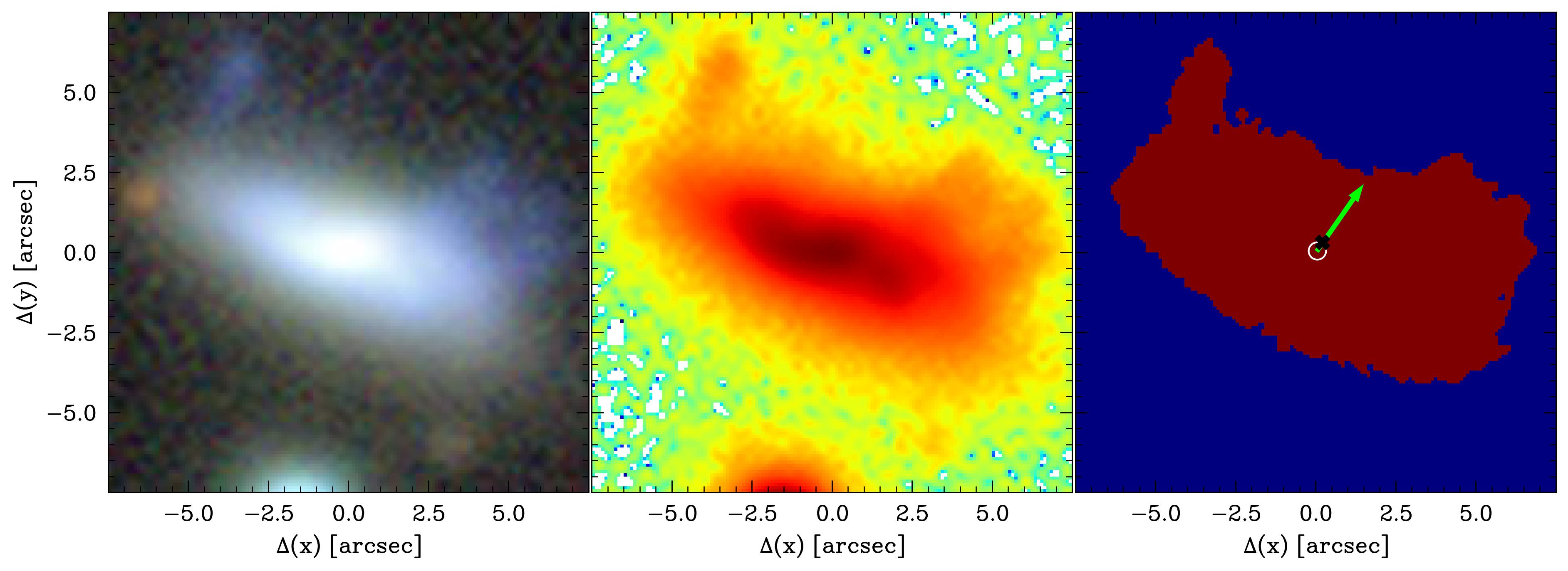}
    \caption{The steps for determining the tail orientation for LS imaging-selected jellyfish galaxy \texttt{C901005481506069}. \textbf{Left panel:} Optical $griz$ composite image from the Legacy Survey \citep{Dey2019}. \textbf{Middle panel:} The corresponding $g$-band image displayed on a logarithmic scale to highlight low-surface-brightness features. \textbf{Right panel:} The binary emission map, where spaxels with confirmed emission detections are assigned a value of 1 (red), and non-detections are 0 (blue). The galaxy centre is indicated by a white circle, while the ionised gas centroid (black cross) is determined according to Equation~\ref{eq:Gas centroid}. The lime arrow shows the tail direction calculated using Equation~\ref{eq:Tail direction}.}
    \label{fig: LS Jellyfish}
\end{figure*}

\end{document}